\documentclass[english,prb,twocolumn]{revtex4}
\usepackage[T1]{fontenc}
\usepackage[latin9]{inputenc}
\usepackage{color}
\usepackage{verbatim}
\usepackage{amsmath}
\usepackage{amssymb}
\usepackage{graphicx}

\makeatletter
\PassOptionsToPackage{caption=false}{subfig} 
\usepackage{hyperref}
\hypersetup{
breaklinks=true,
colorlinks=true,
citecolor=blue,
linkcolor=blue,
filecolor=blue,
urlcolor=blue
}
\IfFileExists{lmodern.sty}{\usepackage{lmodern}}{}

\makeatother

\newcommand{ \Lb }[0]{L}
\usepackage{babel}
\begin{document}

\title{Microwave signatures of $\mathbb{Z}_{2}$ and $\mathbb{Z}_{4}$ fractional Josephson effects
}

\author{P. L. S. Lopes}

\affiliation{Institut Quantique and D\'{e}partement de Physique, Universit\'{e}
de Sherbrooke, Sherbrooke, Qu\'{e}bec, Canada J1K 2R1}

\author{S. Boutin}

\affiliation{Institut Quantique and D\'{e}partement de Physique, Universit\'{e}
de Sherbrooke, Sherbrooke, Qu\'{e}bec, Canada J1K 2R1}

\author{P. Karan}

\affiliation{Institut Quantique and D\'{e}partement de Physique, Universit\'{e}
de Sherbrooke, Sherbrooke, Qu\'{e}bec, Canada J1K 2R1}

\author{U. C. Mendes}

\affiliation{Institut Quantique and D\'{e}partement de Physique, Universit\'{e}
de Sherbrooke, Sherbrooke, Qu\'{e}bec, Canada J1K 2R1}

\author{I. Garate}

\affiliation{Institut Quantique and D\'{e}partement de Physique, Universit\'{e}
de Sherbrooke, Sherbrooke, Qu\'{e}bec, Canada J1K 2R1}
\begin{abstract}
We present a many-body exact diagonalization study of the $\mathbb{Z}_2$ and $\mathbb{Z}_4$ Josephson effects in circuit quantum electrodynamics architectures. 
Numerical simulations are conducted on Kitaev chain Josephson junctions hosting nearest-neighbor Coulomb interactions.
The low-energy effective theory of highly transparent Kitaev chain junctions is shown to be identical to that of junctions created at the edge of a quantum spin-Hall insulator. 
By capacitively coupling the interacting junction to a microwave resonator, we predict signatures of the fractional Josephson effects on the cavity frequency and on time-resolved reflectivity measurements.
\end{abstract}
\maketitle

\section{Introduction}
Josephson junctions (JJs) built at the edges of quantum spin Hall (QSH) insulators have been predicted to display a rich variety of phenomena, which emerge from the interplay between time-reversal (TR) symmetry and the conservation of a local fermion parity. 
In the presence of a dc voltage bias, three theoretical scenarios have been proposed, with distinct periodicities of the Josephson current on the superconducting phase difference across the junction.~\cite{PhysRevLett.75.1831,PhysRevB.95.014505, PhysRevB.79.161408,PhysRevLett.113.036401,PhysRevB.91.081406}

In the first scenario, concerning non-interacting and TR-symmetric JJs, an ac $2\pi$-periodic Josephson effect takes place, together with a dissipative dc current.~\cite{PhysRevLett.75.1831,PhysRevB.95.014505}
This is the ordinary Josephson effect for perfectly transparent weak links.
In the second scenario, involving JJs with broken TR symmetry, the current is dissipationless and its period doubles to $4\pi$. 
Such doubling is the hallmark of hybridized Majorana zero-modes (MZMs) at the edges of the weak link.~\cite{PhysRevB.79.161408}
In the third scenario, entailing TR-symmetric JJs with short-range interactions, the current is non-dissipative and $8\pi$-periodic.~\cite{PhysRevLett.113.036401,PhysRevB.91.081406}
This effect has been attributed to TR-protected $\mathbb{Z}_{4}$ parafermions, fractionalized quasiparticles of conceptual and practical interest.~\cite{PhysRevLett.113.036401} 
The $4\pi$-periodic ($\mathbb{Z}_2$) and $8\pi$-periodic ($\mathbb{Z}_4$) Josephson effects are known as ``fractional'', as opposed to the ``integer'' $2\pi$-periodic Josephson effect.

The experimental realization of fractional Josephson effects constitutes an active research topic in topological condensed matter physics. 
Unexpectedly, recent experiments on QSH JJs have reported Shapiro steps and Josephson radiation consistent with a $4\pi$-periodic Josephson effect,~\cite{PhysRevX.7.021011,Bocquillon_NatNanotech,Wiedenmann_natcomms}
instead of the $2\pi$-periodic or $8\pi$-periodic effects that would have been anticipated for such a TR-symmetric system. 
Consistent explanations for this phenomenon have been put forward in terms of exchange interactions between QSH edge states and nearby charge puddles, which can act as magnetic impurities,~\cite{PhysRevLett.117.267001}
as well as in terms of two-particle inelastic scattering.~\cite{2018arXiv180501137S}

The $8\pi$ Josephson effect remains experimentally elusive to this day.
Its observation requires  weak links of lengths comparable to, or larger than, the superconducting (SC) coherence length.
In addition, a many-body energy gap produced by TR-preserving interactions is needed.
For umklapp interactions, such a gap develops only in the strong coupling limit.~\cite{PhysRevLett.113.036401,PhysRevB.91.081406}
To date, it is unclear whether the condition of strong interactions may be satisfied in real QSH JJs. 
In contrast, spin-flip interactions with magnetic impurities can generate $8\pi$-periodicity both at strong and weak coupling.~\cite{PhysRevLett.117.267001,PhysRevB.95.014505, PhysRevB.96.195421} 
Nevertheless, in the weak coupling regime, interactions with magnetic impurities give a dominant $4\pi$ periodicity.~\cite{PhysRevLett.117.267001}
In addition, for magnetic impurities of spin higher than $1/2$, particularities of the single-ion anisotropies can give rise to $2\pi$ and $4\pi$ periodicities.

In view of the aforementioned challenges, it would be of interest to (i) identify alternative systems where the $8\pi$-periodic Josephson effect can occur, and (ii) develop alternative ways to measure it.
The main objective of the present work is to make theoretical progress along these lines.
Concerning (i),  we establish that the $8\pi$-periodic Josephson effect can take place in JJs built out of Kitaev chains,~\cite{2001PhyU...44..131K} i.e. one-dimensional lattices of spinless fermions with $p$-wave superconductivity. The proposals for physical realizations of Kitaev chains are numerous and under intense experimental investigation (see [\onlinecite{aguado}] and references therein).
Concerning (ii), we propose signatures of the $8\pi$ Josephson effect in circuit quantum electrodynamics (cQED) architectures.

Our study begins in Sec.~\ref{sec:QSH-JJ-EFT}, where we show that a Kitaev chain JJ has the same low-energy effective field theory as the QSH JJ.
This equivalence holds provided that the lattice model is tuned to the regime of a perfectly transparent junction (Sec.~\ref{subsec:Lattice-to-continuum}). In this regime, the lattice model is endowed with an effective low-energy TR symmetry operator  squaring to $-1$, which mimics that of the QSH JJ. Because the low-energy states of the junction are localized within the weak link, finite-sized superconducting electrodes suffice to achieve a good agreement between the continuum and lattice theories (Secs. \ref{subsec:Single-particle-problem} and \ref{subsec:Non-interacting-many-particle-st}). Therefore, we can access physical observables of strongly interacting QSH JJs via exact diagonalization of the Kitaev chain JJ. Specifically, we carry out a lattice analysis of the $\mathbb{Z}_2$ and $\mathbb{Z}_4$ Josephson effects (Secs. \ref{subsec:QSH-meso-JJ}, \ref{subsec:Effective-TR-breaking} and \ref{subsec:TR-Preserving-Interactions}). 
Here, the main advantage over the recent studies of fractional Josephson effects based on bosonization and perturbation theory~\cite{PhysRevLett.113.036401,PhysRevLett.117.267001,PhysRevB.96.195421,PhysRevB.95.014505,PhysRevB.96.165429} is that we have access to the many-body energies \emph{and} wavefunctions, which then allow us to compute physical observables for an arbitrary interaction strength. 

In Sec.~\ref{sec:Circuit-QED}, we apply our theory to determine the influence of strong interactions and quasiparticle fractionalization in cQED measurements of topological JJs.
Recently, cQED architectures~\cite{PhysRevA.69.062320} have been explored, both theoretically and experimentally, as promising venues to probe and characterize topological superconductivity in JJs.~\cite{Cottet:2013ai,PhysRevB.92.245432,PhysRevB.94.115423,2017arXiv171101645H,PhysRevB.88.235401,PhysRevLett.118.126803,PhysRevB.92.134508}
In cQED, a microwave cavity is utilized to monitor, in an efficient and non-invasive way, the discrete energy level dynamics of quantum circuits.~\cite{PhysRevB.92.134508}
Thus far, all cQED studies of topological junctions have neglected the effect of short-range electron-electron interactions.
Accordingly, little is known about the cQED signatures of the $8\pi$-periodic Josephson effect.
By investigating the response of a microwave resonator coupled to a topological JJ (Sec.~\ref{subsec:input-output}), we find that the cavity frequency inherits the anomalous Josephson periodicities and displays a series of kinks and peaks (Sec.~\ref{subsec:pull}) that can be resolved
in the phase-shift of the reflected signal (Sec.~\ref{subsec:shift}). In contrast, the cavity linewidth is unaffected by the presence of the junction, as long as (i) the broadening of the electronic states is small compared to the cavity frequency, and (ii) the cavity frequency is smaller than the energy gaps that protect the fractional Josephson effects (Sec.~\ref{subsec:linewidth}). Finally, Sec.~\ref{sec:Final-Remarks} presents the conclusions, and the appendices contain extra details on the calculations.

\section{Fractional Josephson effects in Kitaev chain junctions\label{sec:QSH-JJ-EFT}}

The objective of this section is to establish an equivalence between the Kitaev chain JJ and the QSH JJ at low energies. We begin by demonstrating that the low-energy continuum expansion of the lattice model exhibits an effective TR symmetry which allows mapping to the QSH JJ.
Then, we proceed with a pedagogical discussion of the fractional Josephson effects that arise when the effective TR symmetry is broken or many-body interactions are turned on. 
After that, many-body spectra and wavefunctions for the $4\pi$- and $8\pi$-periodic scenarios are obtained by exact diagonalization of the lattice model. 
One important conclusion from this section is that the $8\pi$ Josephson effect can occur in Kitaev chains. This statement complements that of Ref.~\onlinecite{PhysRevB.96.165429}, where the authors considered an interacting Rashba nanowire with ``true'' TR symmetry (i.e., no magnetic fields). 
Here, we demonstrate that the $8\pi$-periodic Josephson effect is also possible in Rashba nanowires placed under magnetic fields, because of an effective TR symmetry that emerges at low energies when the JJ has a high transparency. 

\subsection{Lattice and continuum models\label{subsec:Lattice-to-continuum}}

\begin{figure}[t]
\includegraphics[width=0.9\columnwidth]{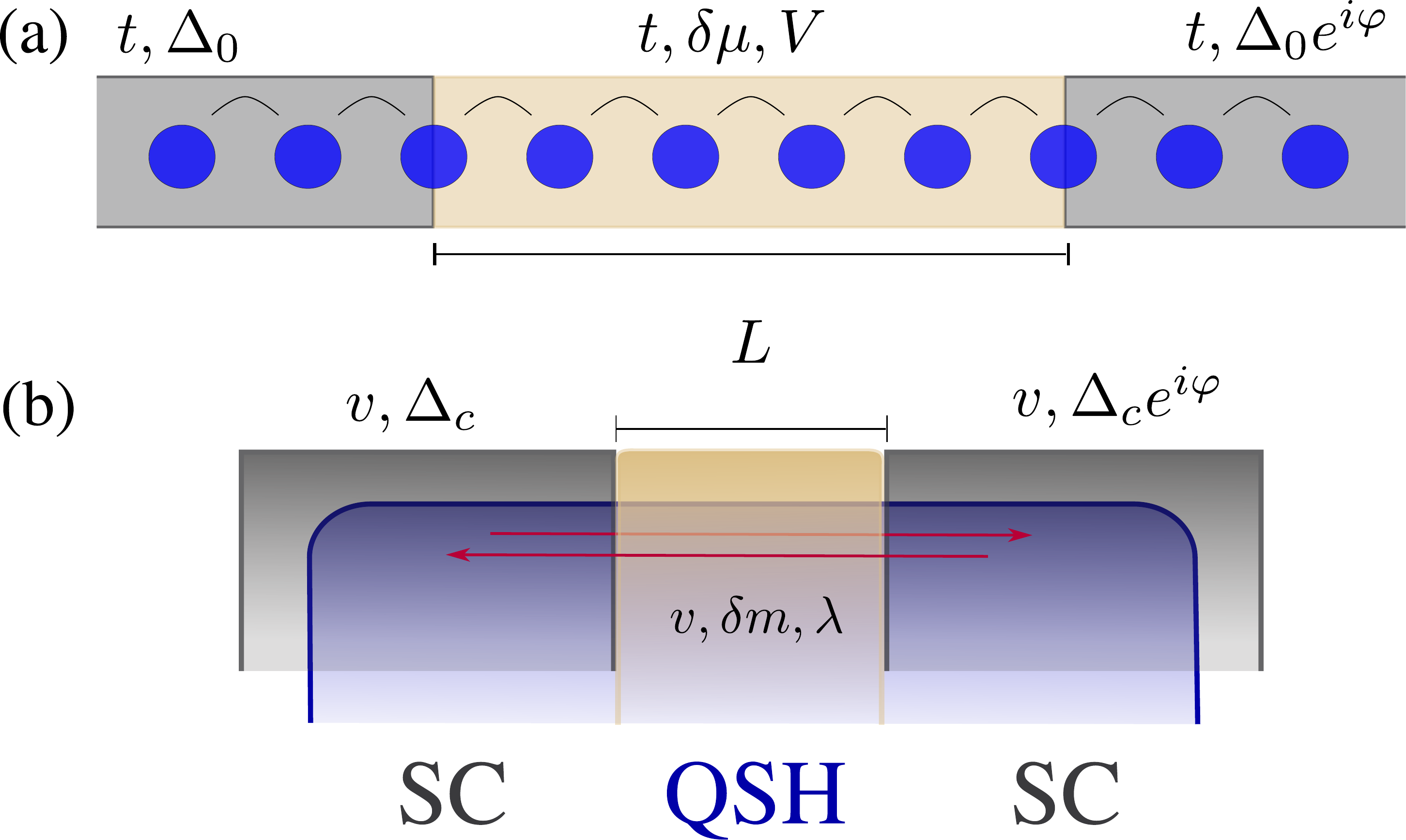}
\caption{(a) Cartoon of a Kitaev chain Josephson junction containing $N$ sites (blue). 
A pair of p-wave superconducting regions (gray) of pairing strength $\Delta_0$ are separated by a normal weak link (yellow) containing $N_L$ sites.
The superconducting phase difference is $\varphi$.
When the hopping amplitude $t$ and the onsite potential for the spinless fermions are uniform throughout the system, 
an effective TR symmetry squaring to $-1$ emerges at low energies. 
Local onsite potentials ($\delta\mu$) break this symmetry, whereas first-neighbor extended
Hubbard interactions ($V$) do not.
(b) Cartoon of a quantum spin-Hall Josephson junction, with a pair of helical edge modes (red arrows) of velocity $v$. At low energies, the Kitaev chain JJ can emulate a QSH JJ. Local onsite potentials and first-neighbour repulsive interactions of the Kitaev JJ map onto magnetic perturbations ($\delta m$) and TR-preserving interactions ($\lambda$) in QSH JJs, respectively. 
\label{fig:QSHJJ}}
\end{figure}

Figure~\ref{fig:QSHJJ}(a) illustrates a Kitaev chain of $N$ sites , whose Hamiltonian reads 
\begin{align}
H_{JJ}\!
&=\!-\sum_l \left[
\left(tc_{l}^{\dagger}c_{l+1}+
	\Delta_{l}
	c_{l}c_{l+1}
	+h.c.
\right)
 +\mu c_{l}^{\dagger}c_{l}\right].
 \label{eq:Hcf} 
\end{align}
Here, $c_{l}$'s are fermion operators at site $l \in \left\{-N/2, \dots, N/2-1\right\}$, $\mu$ is a uniform
chemical potential, $t>0$ is the hopping parameter, and $\Delta_l$ is the complex pairing potential at site $l$. To obtain a Josephson junction with a weak link of length $N_{L}-1$, we consider 

\begin{equation}
\Delta_{l}
=\begin{cases}
\Delta_{0} &, \,l < -N_L/2\\
0 &, \,-N_L/2 \leq l< N_L/2\\
\Delta_{0}e^{i\varphi} &, \,l\geq N_{L}/2
\end{cases},
\end{equation}
where $\varphi$ is the superconducting phase difference across the junction,  and $\Delta_0$ is taken to be real.
For simplicity, we take $N$ and $N_L$ to be even. In this spinless model, the TR operation is simply the complex conjugation $K$.
For the JJs studied in this work, the charging energy is assumed to be much smaller than the Josephson energy and thus $\varphi$ is regarded as a c-number. 

Assuming that the chemical potential is well within the bandwidth ($|\mu|\ll2t$), namely that the chain is well within the topologically nontrivial phase, we can make a low-energy expansion of the fermionic lattice modes close to the two normal-phase Fermi points: $a^{-1/2}c_{l}\approx\left[e^{ik_{F}x}\psi_{R}+e^{-ik_{F}x}\psi_{L}\right]$,
where $a$ is the lattice constant ($x=la$), $\hbar=1$, and $\psi_{R,L}$
are slowly fluctuating right- and left-mover fields. 
The Fermi wavevector $k_{F}$ is defined through $\mu=-2t\cos k_{F}a$. 
To leading order in a gradient expansion of $a$,
and neglecting fast oscillating terms, 
Eq.~(\ref{eq:Hcf}) becomes

\begin{align}
 \label{eq:contH}
H_{JJ}\left(\varphi\right) & \approx v\int\! dx\left(\psi_{R}^{\dagger}\left(-i\partial_{x}\right)\psi_{R}-\psi_{L}^{\dagger}\left(-i\partial_{x}\right)\psi_{L}\right) \\
 &
 +\int \!\!dx\left(\Delta_c \Theta(|x|-L/2) e^{i \Theta(x) \varphi}\psi_{L}\psi_{R}+h.c.\right)
 \nonumber
 ,
\end{align}
where $\Theta(x)$ is the step function, $v=2at\sin\left(k_{F}a\right)$ is a velocity, $\Delta_c=2 \Delta_0 \sin\left(k_{F}a\right)$ is the effective pairing potential, $L=(N_L-1)a$ is the length of the weak link, and the superconducting phase was globally shifted by $\pi/2$. As illustrated in Fig.~\ref{fig:QSHJJ}(b),  the same Hamiltonian describes a JJ at the edge of a spin-momentum-locked QSH insulator with proximitized
s-wave superconductivity.~\cite{PhysRevB.79.161408,PhysRevLett.113.036401} Next, we consider 
possible antiunitary TR operators, which commute with our low-energy description of $H_{JJ}(\varphi)$.

The lattice level TR operator $\mathcal{T}_+$ acts on the continuum basis 
by exchanging $L$ and $R$ modes up to a gauge-dependent phase. For Eq.~\eqref{eq:contH}, TR acts on the operators as 
$\mathcal{T}_+ \psi_L \mathcal{T}_+^{-1} = i \psi_R$,
$\mathcal{T}_+ \psi_R \mathcal{T}_+^{-1} = i \psi_L$, and the lattice level symmetry is preserved such that 
$[H_{JJ}(n\pi)\,,\,\mathcal{T}_+] =0$, for $n\in\mathbb{Z}$. Defining a spinor $\left(\psi_{R},\,\psi_{L}\right)$ with the left-
and right-moving modes, and a set of Pauli matrices $\tau_{i}\, (i=x,y,\,z$)
acting on this space, the so-called \textit{first-quantized} description of this TR operator is $T_+ = i \tau_x K$, with $T_+^2 = +1$. In addition, we can define a second antiunitary operator $\mathcal{T}_-$ which also commutes with Eq.~\eqref{eq:contH} at $\varphi = n\pi$ ($n \in \mathbb{Z})$ and with first-quantized representation $T_-= i \tau_y K$. Since $T_-^2 = -1$, this additional symmetry enforces Kramers degeneracies at TR-invariant superconducting phase differences.

Even though the fermions $\psi_{R}$ and $\psi_{L}$ carry no spin degrees of freedom, their Hamiltonian
displays the same symmetries and behavior as that of a QSH edge state. Unlike in the case of the QSH edge, however, this $T_-=i\tau_{y}K$ TR symmetry is only effective. 
First, it crucially relies on the validity of neglecting the fast oscillating terms in the low-energy expansion leading to Eq.~\eqref{eq:contH}. In order to be valid, this approximation requires the superconducting coherence length $\xi_0=\hbar v/\Delta_{c}=ta/\Delta_{0}$ to obey $\xi_0 \gg 2\pi/k_F$, which will be satisfied for lattice parameters such that $t \gg \Delta_0 $ ($\xi_0 \gg a$). Second, certain perturbations of the lattice Hamiltonian (\ref{eq:Hcf}), such as local spatial inhomogeneities in the hopping parameter or in the chemical potential, produce terms in the continuum approximation that do not commute with $\mathcal{T}_-$, leading to single-body backscattering terms between left- and right-movers.
In short, $\mathcal{T}_-$ is a low-energy symmetry of the Kitaev chain JJ only when the transparency of the junction is unity.
Extended Hubbard interactions -- the simplest two-body terms in the Kitaev chain -- preserve $\mathcal{T}_-$ at low energies. 
For now, we proceed with the non-interacting and fine-tuned TR-preserving scenario.

\subsection{Single-particle states\label{subsec:Single-particle-problem}}

In this subsection, we validate the continuum expansion of the lattice model by calculating and comparing the spectra of Eqs.~(\ref{eq:Hcf}) and~(\ref{eq:contH}).
This exercise will set the notation for the following sections. Since Eq.~(\ref{eq:contH}) has been previously solved,~\cite{PhysRevB.79.161408,PhysRevLett.113.036401,PhysRevB.95.014505,PhysRevB.95.235134} here we review the main results rapidly but pause on some intricacies that are rarely discussed in the literature. 

Measuring energies and lengths in units of $\Delta_{c}$ and $\xi_{0}$ respectively, Eq.~(\ref{eq:contH}) can be recast in the
Bogoliubov de Gennes (BdG) form
\begin{align}
H_{JJ}\left(\varphi\right) & \approx\frac{1}{2}\int \! dx\,\Psi^{\dagger}h\left(\varphi\right)\Psi,
\end{align}
where $\Psi=\left(\psi_{R},\psi_{L},\psi_{L}^{\dagger},-\psi_{R}^{\dagger}\right)^{T}$.
In this basis, the single-particle BdG Hamiltonian reads
\begin{align}
h\left(\varphi\right) &=\left[-i\tau_{z}\partial_{x}\right]\rho_{z}+\Theta(|x|-\Lb/2)\rho_{x}e^{i\rho_{z} 
\Theta(x)\varphi},
\end{align}
where $L/\xi_0 \rightarrow \Lb$, Pauli matrices $\rho_{i}$ act in the Nambu particle-hole space, and,
as mentioned above, $\tau_i$ matrices act in the $\psi_{L},\psi_{R}$ space. 

Using $\left[h,\tau_{z}\right]=0$, we decompose the Hilbert space
in two $\tau_{z}$ eigensectors and solve
\begin{equation}
h_{\tau}\left(\varphi\right)\psi_{\tau}=E_{\tau}\psi_{\tau}\label{eq:sp_energ}
\end{equation}
with wavefunctions of the form $\psi_{+}=\left(u_{+},0,v_{+},0\right)^{T}$ and
$\psi_{-}=\left(0,u_{-},0,v_{-}\right)^{T}$ obeying continuous boundary
conditions at $x=\pm \Lb/2$. 

\subsubsection{Energy spectrum of Eq.~(\ref{eq:sp_energ})}

Let us concentrate on the Andreev bound state
(ABS) spectrum, i.e. states with discrete energies inside the bulk SC gap ($|E_{\tau}|<1$). The ABS energies are determined by the solutions of the transcendental equation

\begin{equation}
\tan(\tau \Lb E_{\tau})=\frac{\sqrt{1-E_{\tau}^{2}}-\tau E_{\tau}\tan\left(\frac{\varphi}{2}\right)}{\tau E_{\tau}+\sqrt{1-E_{\tau}^{2}}\tan\left(\frac{\varphi}{2}\right)},\label{eq:trans_SB}
\end{equation}
which reduces to an earlier result~\cite{PhysRevLett.113.036401} at $\varphi=\pi$. For each value of $\varphi$, the solutions $E_{n,\tau}\left(\varphi\right)$ are discrete and labelled with the indices $n$ and $\tau$. The latter index caracterizes the slope of the energy eigenvalue as a function of $\varphi$: $\tau=+1$ for negative slope, $\tau=-1$ for positive slope.

\begin{figure}[t]
\includegraphics[width=0.9\columnwidth]{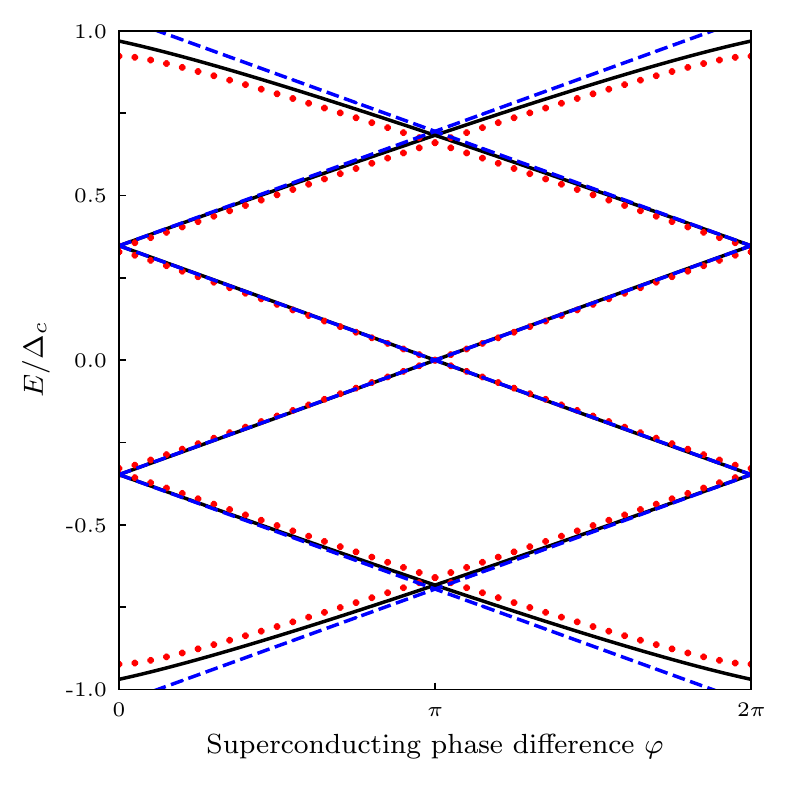}
\caption{Non-interacting QSH JJ single-particle spectrum obtained for $\mu=0$, $\Delta_{0}=t/2$
($\xi_0 = 2 a$). Red dots are obtained from numerical diagonalization of the Kitaev
chain JJ ($N_L=8$, $N=200$), black full curves are obtained from solving the
continuum effective theory (cf. Eq.~(\ref{eq:trans_SB}), with $L=3.5\xi_0$). 
No adjustments of parameters are made. Blue dashed
curves are obtained by an effective model where the superconducting
banks of the junction are substituted by point-like leads and an effective
pairing 
$\Delta_{\mathrm{eff}}\approx0.77\Delta_{0}$.\label{fig:analytics_vs_numerics_SB}}
\end{figure}

Figure~\ref{fig:analytics_vs_numerics_SB} displays the solutions of
Eq.~\eqref{eq:trans_SB} (full black) and a single-particle diagonalization
of Eq.~(\ref{eq:Hcf}) (red dots). 
A good agreement is obtained between the two sets of curves for energies well inside the bulk SC gap.
The agreement can be made even better by increasing the ratio $\xi_0/a$, which further suppresses the $\mathcal{T}_-$-breaking fast oscillating terms.

The structure of the energy eigenvalues in Fig.~\ref{fig:analytics_vs_numerics_SB} is constrained by the Nambu particle-hole operator $C=\rho_{y}\tau_{y}K$ and the TR operator $T_-=i\tau_{y}K$, which impose
\begin{align}
\tau_{y}\rho_{y}h\left(\varphi\right)\rho_{y}\tau_{y} & =-h^*
\left(\varphi\right)
\nonumber \\
\tau_{y}h\left(\varphi\right)\tau_{y} & =
h^*
\left(-\varphi\right).
\end{align}
These relations in turn enforce 
\begin{align}
E_{\tau}\left(\varphi\right) & =-E_{-\tau}\left(\varphi\right)\nonumber \\
E_{\tau}\left(\varphi\right) & =E_{-\tau}\left(-\varphi\right),
\end{align}
where the left- and right-hand-sides need not correspond to the same value of $n$.
As a consequence of TR symmetry and the $2\pi$-periodicity of the Hamiltonian, 
different ABS cross at $\varphi=\pi$ (or multiples thereof). The index $n$ can be used to identify these crossings, with $n>0$ if the crossing happens
at positive energies, $n=0$ if the crossing is at vanishing energy,
$n<0$ for crossings at negative energies. 
The number $M$ of positive-energy ABS crossings at $\varphi=\pi$ depends on the length of the weak link
and is fixed by $-\frac{\pi}{2}<\Lb-M\pi\leq\frac{\pi}{2}$,
as can be concluded from Eq.~(\ref{eq:trans_SB}). 
With the parameter values of Fig.~\ref{fig:analytics_vs_numerics_SB} we have  $M=1$, which is the minimum necessary for the $\mathbb{Z}_4$ Josephson effect to be discussed below. 
For concreteness, we will keep this number of crossings for the remainder of this paper. 

To gain some analytical insight the characteristic energy scale of the ABS modes, we take an approach of replacing the SC electrodes by point-like SC leads. In this case, following the standard procedure of effective field theories, we fit for an effective SC pairing strength that returns
the correct energy spectrum (see Ref.~{[}\onlinecite{Kaplan_lects}{]} for an illuminating discussion). 
After putting back the units, this exercise results in 
~\cite{PhysRevB.95.235134} 
\begin{equation}
E_{n,\tau}=
\frac{\Delta_{\mathrm{eff} }\xi_0}{2L}
\left[\pi\left(2n+\tau\right)-\tau\varphi\right],
\end{equation}
thereby uncovering the scaling of ABS energies with the Thouless energy $E_{T}=\Delta_{c} \xi_{0} /L=v/L$.
Fitting an effective SC pairing of 
$\Delta_{\mathrm{eff}}\approx0.77\Delta_{c}$, one recovers the blue dashed curves in Fig.~\ref{fig:analytics_vs_numerics_SB},
which again match the numerical data at low energies, as expected for an effective field theory.

\subsubsection{Wavefunctions of Eq.~(\ref{eq:sp_energ})}

The nonzero components of the $\psi_{n>0,\tau}$ bound states are~\cite{PhysRevLett.113.036401,PhysRevB.95.014505}
\begin{align}
u_{n,\tau} 
& =A_{n,\tau}e^{-\sqrt{1-E_{n,\tau}^{2}}\left|x-l\left(x\right)\right|}\left(-1\right)^{n}e^{i\tau E_{n,\tau}l\left(x\right)}
\\
v_{n,\tau} 
& = -\tau A_{n,\tau}e^{-\sqrt{1-E_{n,\tau}^{2}}\left|x-l\left(x\right)\right|} e^{i\left(\frac{\varphi}{2}\right)}e^{-i\tau E_{n,\tau}l\left(x\right)},
\nonumber 
\label{eq:wavefs}
\end{align}
where 
\begin{equation}
l\left(x\right)=\begin{cases}
x & \text{if}\,\left|x\right|<\Lb/2\\
\text{sgn}\left(x\right)\frac{\Lb}{2} & \text{if}\,\left|x\right|>\Lb/2
\end{cases}
\end{equation}
and the normalization factor reads 
$
\left|A_{n,\tau}\right|
=\left\{2\left[\Lb+(1-E_{n,\tau}^{2})^{-1/2}\right]\right\}^{-1/2}
$
. 
To access the $\psi_{n<0,\tau}$ states, it suffices to apply the Nambu particle-hole transformation $C=\rho_{y}\tau_{y}K$. The eigenstates obey the orthogonality relations
\begin{equation}
\int dx\psi_{n,\tau}^{\dagger}(x, \varphi)\psi_{n',\tau}(x, \varphi)=\delta_{\tau,\tau'}\delta_{n,n'}
\end{equation} and, if supplemented with the continuum of scattering states, the completeness relation

\begin{equation}
\sum_{n,\tau}\psi_{n,\tau}\left(x, \varphi\right)\otimes\psi_{n,\tau}^{\dagger}\left(x', \varphi\right)=\mathbb{I}_{4\times4}\delta\left(x-x'\right)
\end{equation}
is respected, where $\mathbb{I}_{4\times4}$ is the $4\times4$ identity matrix.

For later discussion on the transformation properties of the many-body states under TR, it is convenient to consider the action
of $\mathcal{T}_-$ on the above wavefunctions.
The transformation rules may be written as
\begin{align}
\label{eq:sb_TR}
i\tau_{y}\psi_{n,\tau}^{*}\left(x,\varphi\right)&=\tau\psi_{n+\tau,-\tau}\left(x,-\varphi\right)\nonumber\\
&=-\tau\psi_{n,-\tau}\left(x,2\pi-\varphi\right).
\end{align}

\subsection{Non-interacting many-particle states\label{subsec:Non-interacting-many-particle-st}}

The single-particle wavefunctions and energies from the previous subsection allow us to construct non-interacting many-particle states in the continuum approximation. This construction will be useful for later discussion on interacting JJs. 
The starting point is to expand the field operators in terms of ABS as 
\begin{align}
\Psi\left(x\right) & =\sum_{n,\tau}\psi_{n,\tau}\left(x,\varphi\right)a_{n,\tau}\left(\varphi\right)\nonumber \\
a_{n,\tau}\left(\varphi\right) & =\int dx\psi_{n,\tau}^{\dagger}\left(x,\varphi\right)\Psi\left(x\right),\label{eq:ABS_modes}
\end{align}
where the operator $a_{n,\tau}$ annihilates the ABS labeled with $(n,\tau)$.
For the junction length $L\approx\pi\xi_0$
chosen above, we may
limit ourselves to the six lowest-energy states, 

\begin{align}
\left|j;\varphi\right\rangle ,\  & j=0,...,5,\label{eq:MB_states}
\end{align}
which are plotted in Fig.~\ref{fig:MB_no_int} and presented in more detail in Appendix \ref{subsec:Single-particle-many-body-states}.
These states are built from fixing $\left|0;\varphi=0\right\rangle $ with all negative energy single-particle states
filled. The excitations over the ground state involve ``particle-hole'' pairs composed of positive energy quasiparticles, together with their Nambu conjugate quasiholes. The total number of BdG quasiparticles is the same in all states.

\begin{figure}[t]
\includegraphics[width=0.9\columnwidth]{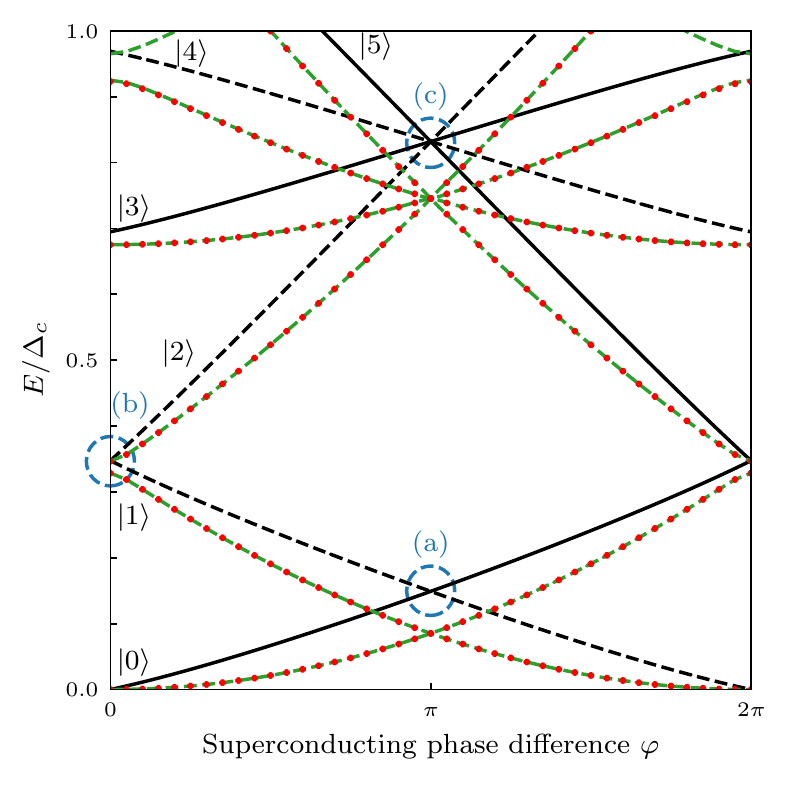}
\caption{Many-body energy spectrum in the absence of interactions. 
The parameter values are the same as in Fig.~\ref{fig:analytics_vs_numerics_SB}. Black curves are obtained by solving the transcendental equation (\ref{eq:trans_SB})
and by thereafter building many-body energies from Eq.~(\ref{eq:MB_states}); see Appendix~\ref{subsec:Single-particle-many-body-states} for further details.
Full versus dashed black lines correspond
to different eigenvalues of the parity operator (\ref{eq:parity}).
Contributions from single-particle scattering states are ignored,
which generates a mismatch with the remaining data. 
Red dots correspond to a single-body exact diagonalization of Eq.~(\ref{eq:Hcf}) for
chains of $N=200$ sites, with many-body energies built in a similar
way as the black lines from Eq.~(\ref{eq:MB_states}), but scattering
states taken into account. 
Green dashes are also obtained from the
lattice Hamiltonian, but from a brute force diagonalization in the
many-body Fock space for a chain with $N=26$ sites. Avoided crossings labeled (b) are due to $\mathcal{T}_-$-breaking terms in the lattice Hamiltonian which become negligible in the $\xi_0 \gg a$ limit (see Fig.~\ref{fig:MB_latt_vs_EFT} of Appendix~\ref{subsec:Single-particle-many-body-states}).
\label{fig:MB_no_int}}
\end{figure}

With the states in Eq.~(\ref{eq:MB_states}) and the single-particle energies from the previous subsection, one can build the low-energy many-body spectrum. Alternatively, one can perform a brute-force exact diagonalization of the lattice Hamiltonian in the full many-body Fock space, without any reference to single-particle states. The results of both approaches are depicted in Fig.~\ref{fig:MB_no_int}. 

On the one hand, we find an excellent agreement between the single-body lattice diagonalization (red dots), where we use $N=200$ sites, and the many-body numerics (green dashes), where we use $N=26$. 
This indicates that the finite size effects originating from the superconducting leads are not significant.
On the other hand, only a fair agreement is obtained between the continuum (black curves) and the lattice numerics. This discrepancy can be tracked down to the $\varphi$-dependent contribution of the continuum of scattering states to the ground state energy,~\cite{beenakker_QPC} 
which is not captured in the continuum analysis. A better comparison between the effective theory and the exact diagonalization of the lattice model can be achieved by subtracting off the ground state energy at each value of $\varphi$ separately. The obtained excitations energies are physically observable e.g. in the tunneling density of states~\cite{PhysRevLett.113.036401} or in the cQED response functions evaluated in the next section. 
When comparing these excitation energies, a much improved agreement is found between the exact diagonalization results and the analytical results (see Fig.~\ref{fig:MB_latt_vs_EFT} in Appendix A).

On a related note, the avoided crossings at $\varphi=0,\,2\pi$ between the states $|1\rangle$ and $|2\rangle$ (cf. the green dashes and red dots in Fig.~\ref{fig:MB_no_int})
result from the finite ratio of $\xi_0/a$ leading to the continuum Hamiltonian being only approximately equivalent to the lattice model, with small fast-oscillating $\mathcal{T}_-$-breaking terms lifting Kramers degeneracy. We have verified that these anticrossings are eliminated by increasing
the ratio of $\xi_{0}/a$, which is easily done for single-particle
diagonalizations, but not for the many-body case (due to system size
limitations).

For the remainder of this subsection, we study the rich structure of crossings in Fig.~\ref{fig:MB_no_int} from a symmetry point of view. We begin by recalling that TR acting in the second-quantized operators yields~\cite{Shinsei}
\begin{align}
\label{eq:TRsim}
{\cal T}_-\Psi\left(x\right){\cal T}_-^{-1} & =i\tau_{y}\Psi\left(x\right).
\end{align}
Combining Eqs.~(\ref{eq:sb_TR}), (\ref{eq:TRsim})and (\ref{eq:ABS_modes}), we get
\begin{align}
{\cal T}_-a_{n,\tau}\left(\varphi\right){\cal T}_-^{-1} & =\tau a_{n+\tau,-\tau}\left(-\varphi\right)\nonumber \\
 & =-\tau a_{n,-\tau}\left(2\pi-\varphi\right).
\end{align}
Consequently, the action of ${\cal T}_-$ on the many-body states of Eq.~(\ref{eq:mbstates}) at $\varphi=0$ returns (up to phase factors) 
\begin{align}
{\cal T}_-\left|0;0\right\rangle  & \sim\left|0;0\right\rangle \nonumber \\
{\cal T}_-\left|1;0\right\rangle  & \sim\left|2;0\right\rangle  \\
{\cal T}_-\left|3;0\right\rangle  & \sim\left|3;0\right\rangle ,\nonumber
\end{align}
with the other states being either invariant or having partners at
higher energies. At $\varphi=\pi$, one gets
\begin{align}
{\cal T}_-\left|0;\pi\right\rangle  & \sim\left|1;\pi\right\rangle \nonumber \\
{\cal T}_-\left|2;\pi\right\rangle  & \sim\left|5;\pi\right\rangle \label{eq:TR_pi} \\
{\cal T}_-\left|3;\pi\right\rangle  & \sim\left|4;\pi\right\rangle . \nonumber
\end{align}
These transformations demonstrate that many of the crossings in the spectrum of Fig.~\ref{fig:MB_no_int} are
protected by the effective TR invariance of the low-energy physics
(or the true TR invariance of the QSH edge modes, in the case of a QSH JJ). 
Yet, some of the crossings therein are protected by another symmetry as well, namely the local fermion parity. The local fermion-parity operator counts the parity of the number of ABS excitations in the many-body state. It can be written as
\begin{equation}
P_{\text{in}}\left(\varphi\right)\equiv\left(-1\right)^{\left(a_{0,+}^{\dagger}a_{0,+}+\sum_{n>0,\tau}a_{n,\tau}^{\dagger}a_{n,\tau}\right)},\label{eq:parity}
\end{equation}
where the sum over $n$ is done among the discrete-energy bound-states only ($n=1$ only, for our parameter values). 
\footnote{This definition of the parity operator is not unique. The modes contributing to it must be spatially localized in the weak-link and the number of such modes depends on the length of the link. Also, in the absence of TR breaking perturbations, the connection of the ABS modes with the continuum of scattering dictates that a cutoff must be introduced, in a rather arbitrary way, in the mode sum, depending on $\varphi$. Finally, the definition is sensitive to the choice of the reference ground state. The form displayed in this work is in accordance with our conventions, but may be straightforwardly adjusted to other conventions.} An application of this operator over the many-body states written explicitly in Eq.~(\ref{eq:mbstates}) returns the pattern of full and dashed black curves
displayed in Fig.~\ref{fig:MB_no_int}. Importantly, $P_{\rm in}$ is conserved at every $\varphi$  as long as the total fermion parity of the system is conserved.

The conservation of $P_{\rm in}$ and TR allows to understand the various crossings in Fig.~\ref{fig:MB_no_int}. At $\varphi=0$, $\left|1;0\right\rangle $ and
$\left|2;0\right\rangle $  have the same parity and are related
by TR symmetry; they constitute Kramers partners. 
At $\varphi=\pi$, TR operation
connects states of opposite parity.
Accordingly, the degeneracy between $\left|0;\pi\right\rangle $ and $\left|1;\pi\right\rangle$ (two states of opposite parity) is protected by both TR symmetry and the conservation of $P_{\rm in}$.
Indeed, in the topological phase, the parity eigenvalues of the two lowest-energy many-body states are inverted when going from $\varphi=0$ to $\varphi=2\pi$, which requires a band crossing in between.  

The higher energy 4-fold crossing at $\varphi=\pi$ is only partly protected. 
On the one hand, the degeneracy between $\left|2;\pi\right\rangle$ and $\left|5;\pi\right\rangle$, as well as the degeneracy between $\left|3;\pi\right\rangle$ and $\left|4;\pi\right\rangle$ are enforced by both TR symmetry and the conservation of $P_{\rm in}$. On the other hand, the degeneracy between $\left|2;\pi\right\rangle$  and $\left|3;\pi\right\rangle$ is ``accidental'' and guaranteed only at the non-interacting level. Indeed, we will show below that TR- and parity-preserving interactions introduced at the lattice level break the 4-fold degeneracy into a pair of 2-fold crossings,
as previously proposed in the context of QSH JJ.~\cite{PhysRevLett.75.1831,PhysRevB.95.014505, PhysRevB.79.161408,PhysRevLett.113.036401,PhysRevB.91.081406} 

\subsection{Fractional Josephson effects: phenomenology \label{subsec:QSH-meso-JJ}}

Having understood the low-energy spectrum of the non-interacting junction, it is useful to embark on a pedagogical discussion of the different fractional Josephson effects listed in the Introduction.
The different effects can be distinguished by focusing on the several crossings that take place in the energy spectrum of Fig.~\ref{fig:MB_no_int}, at $\varphi= n \pi$ ($n\in\mathbb{Z}$). To guide the explanation, we use the labels ``(a)'' for the lowest 2-fold crossing at $\varphi=\pi$, ``(b)'' for the lowest 2-fold crossings at $\varphi=0$ and ``(c)'' for the 4-fold crossing at $\varphi=\pi$. 
 
When all (a), (b) and (c) crossings are preserved (like in Fig.~\ref{fig:MB_no_int}), 
the ABS energy levels are continuously connected, as a function of $\varphi$, with the continuum of scattering states of energies greater than $\Delta_{c}$. As a consequence, dc-voltage biasing the junction leads to a time-dependent evolution of the states that eventually connects the ground state with the continuum of scattering states, thereby generating a dissipative dc contribution on top of a $2\pi$-periodic Josephson current.~\cite{PhysRevLett.75.1831,PhysRevB.95.014505} In order to have only the purely ac component of the Josephson current, it is necessary to disconnect the ABS from the continuum by opening a gap either at (a), (b)
or (c).

The crossing at (a) is gapped in topologically trivial JJs, which lack MZMs and do not have the corresponding conserved local-fermion-parity. Accordingly, the lowest curve in Fig.~\ref{fig:MB_no_int} fully separates from the rest. The evolution of this state as function of $\varphi$ is $2\pi$-periodic, corresponding to a standard, dissipationless, Josephson effect.

A second possible scenario involves lifting the crossing (b). 
Since this crossing is protected by TR symmetry alone, it can be gapped
by applying a magnetic perturbation on the QSH JJ,~\cite{PhysRevB.79.161408} or by a adding potential barrier (which breaks $\mathcal{T}_-$ symmetry) in the Kitaev chain JJ. The gap scales with the strength of the TR-breaking perturbation, which is responsible for localizing the Majorana modes at the boundaries of the weak link.
Due to this gap,  a doublet of states (the crossing at (a) being
protected by $P_{\rm in}$ conservation) becomes disconnected from
the remaining states, including the scattering ones, and the $4\pi$-
periodic fractional Josephson effect arises. This effect is characteristic of weakly hybridized MZMs allowing for single-electron tunneling through the junction. 

The third and last scenario arises from lifting the 4-fold degeneracy at (c). This crossing, composed of states with one- and two-quasiparticle excitations, exists only if the JJ can accommodate at least three discrete ABS levels with energies
smaller than $\Delta_{c}$, cf. Fig.~\ref{fig:analytics_vs_numerics_SB}. TR- and $P_{\rm in}$-conserving interactions can lift this 4-fold crossing in two pairs of TR- and parity-protected crossings. In a QSH JJ, umklapp interactions (at half-filling) or interactions with magnetic impurities (at any filling) are known to lift the 4-fold degeneracy.~\cite{PhysRevLett.113.036401,PhysRevB.96.195421,PhysRevLett.117.267001,PhysRevB.95.014505,PhysRevB.91.081406,PhysRevB.96.165429}
The case of an interacting Kitaev chain JJ will be discussed below. The consequence of the gap opening at (c) is an $8\pi$-periodic fractional Josephson effect characterized by transport
of charges $e/2$ through the junction.

\begin{figure}[tb]
\includegraphics[width=1\columnwidth]{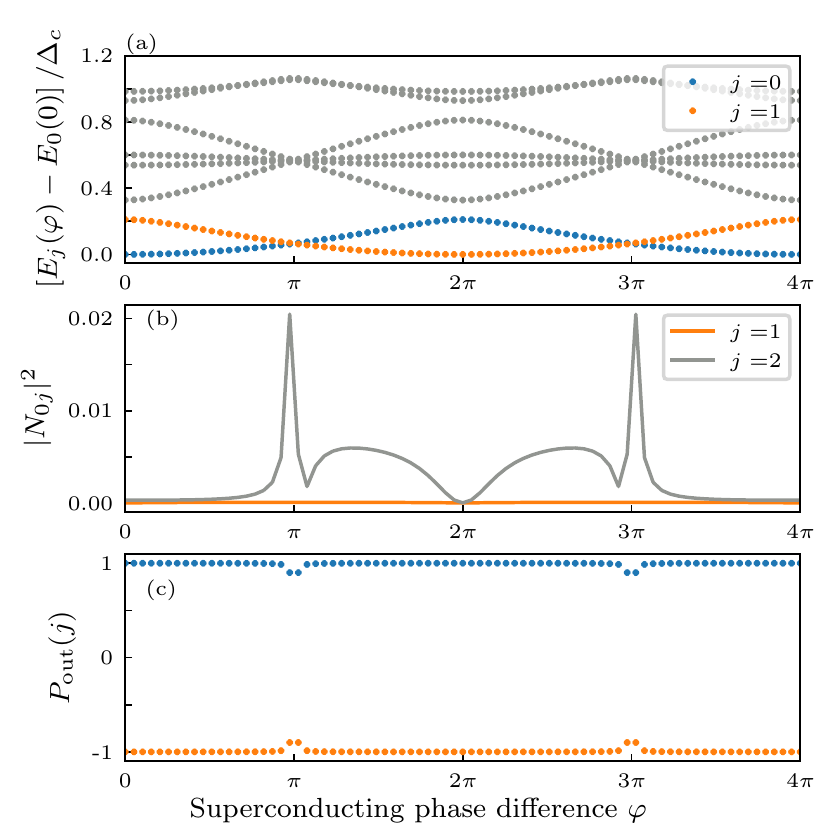}
\caption{Exact diagonalization of a non-interacting Kitaev chain JJ with broken effective TR. 
The parameter values are $\mu=0$, $\Delta_{0}=t/2$,
$N_{L}=8$, $N=26$, $\delta\mu/t=1.3$. 
(a) Many-body spectrum. The blue and orange curves form the ground state multiplet. 
Grey states are excited states. (b)  Off-diagonal matrix elements of the total number operator $\hat{N}$ for the lowest many-body states.  The vanishing of $N_{01}\left(\varphi\right)$ follows the conservation of the local fermion-parity in panel (c). 
(c) The parity of occupation of the non-local fermion in Eq.~\eqref{eq:Pout} for the states forming the ground state multiplet.
Parity conservation protects the crossings in the ground state doublet and enforces the $4\pi$ periodicity of the blue and orange states in panel (a).  
\label{fig:TRB_spec}}
\end{figure}

\subsection{T-breaking perturbations\label{subsec:Effective-TR-breaking}}

In this subsection, we consider a junction where the effective TR symmetry $\mathcal{T}_-$ is broken and the $4\pi-$periodic Josephson effect arises. This can be achieved in several ways but, to keep the analogy with the QSH JJ interrupted by a magnetic insulator, we choose to implement
a potential barrier inside the junction,
\begin{equation}
H_{SB}=\delta\mu\sum_{l>-N_L/2}^{N_L/2-2}c_{l}^{\dagger}c_{l}.\label{eq:HSIS}
\end{equation}
For more appreciable effects, we take the perturbation to be larger than the bandwidth ($\delta\mu>t$). This term transforms the normal weak link into a trivial insulator and thereby localizes Majorana modes at the edges of each SC bank, which hybridize perturbatively. 
From the point of view of our continuum theory, $H_{SB}$ leads to single-body backscattering between the left- and right-movers, violating the effective TR symmetry.

In order to make a seamless connection with the interacting case discussed in the next subsection, we perform an exact diagonalization of the non-interacting TR-broken junction in the  many-body Fock space. As the lattice Hamiltonian preserves the total fermion parity, we project the Fock space into the subspace of an even total number of fermions. As seen in Fig.~\ref{fig:TRB_spec}(a), the subgap part of the spectrum is disconnected from the scattering states, leaving the blue and orange bands as the ground state doublet. As expected, the avoided crossing happens at the crossing (b) of the spectrum of Fig.~\ref{fig:MB_no_int}.

\begin{figure}[tb]
\includegraphics[width=1\columnwidth]{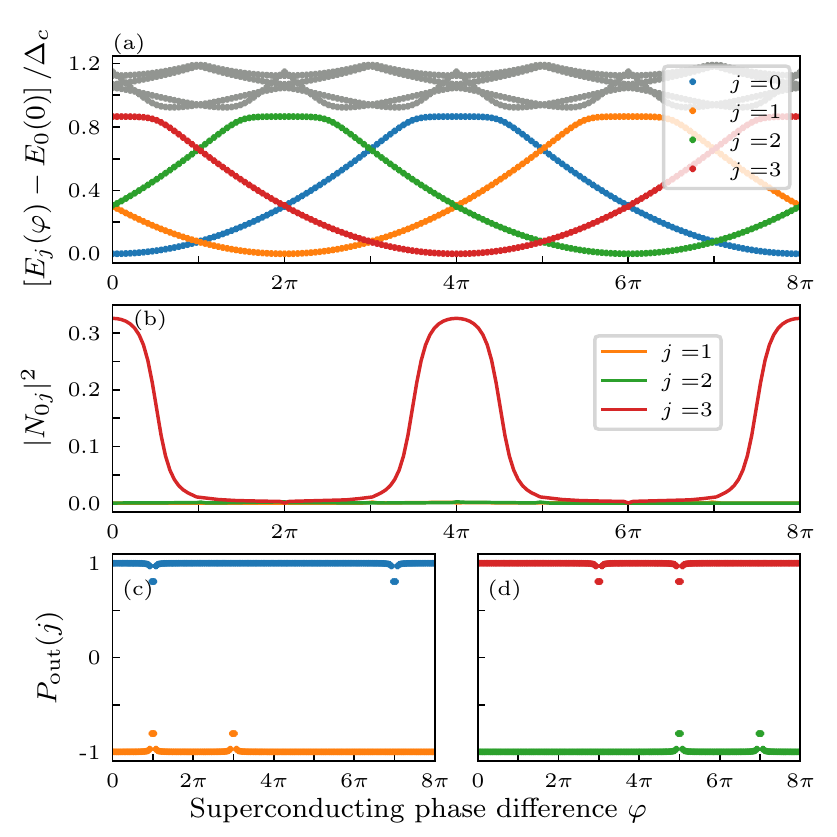}
\caption{Exact diagonalization of the interacting Kitaev chain JJ.
The parameter values are  $\mu=0$, 
$\Delta_{0}=t/2$, 
$N_{L}=8$, $V=2t$ and $N=26$.
(a) Many-body spectrum. Colored states form the 4-fold ground state multiplet. 
Grey states are excited states. (b)  Off-diagonal matrix elements of the total number operator $\hat{N}$ between the $i=0$ (blue) band and the rest of the states forming the ground state multiplet. The matrix elements between states of opposite local fermion parity vanish. The matrix elements between states of the same parity are nonzero, except when they cross. At the crossing points, the effective TR symmetry enforces the vanishing of the matrix element of $\hat{N}$. (c) Parity of occupation of the non-local fermion in Eq.~\eqref{eq:Pout} for the states forming the ground state multiplet.
\label{fig:MB_spec}} 
\end{figure} The continuation of the colouring through the crossings at $\varphi=\pi$ and $3\pi$ in Fig.~\ref{fig:TRB_spec}(a) is justified by the the conservation of $P_{\rm in}$ (cf. Eq.~(\ref{eq:parity})). 
This form of the parity operator cannot be easily accessed from the many-body exact diagonalization, which circumvents the single-particle energy levels. 

To verify the protection of the crossings, we consider instead the parity of occupation of a non-local state built out of the MZMs $\Gamma_L$  and $\Gamma_R(\varphi)$ located respectively at the left and right {\em outer} ends of the chain.~\cite{2001PhyU...44..131K} 
Since the parity of the total number of fermions has been fixed, the parity of this non-local outer mode, 
with operator corresponding to
\footnote{
In the particular case of a perfectly dimerized fine-tuned Kitaev chain ($\mu=0$, $\Delta_0=t$), MZMs are localized on single sites and the parity operator is simply $P_{\text{out}}=(c_{N/2-1}e^{i\varphi/2}+c_{N/2-1}^{\dagger}e^{-i\varphi/2})(c_{-N/2}^{\dagger}-c_{-N/2})$. Away from this point, MZMs extend over several lattice sites (see eq. (14) from Ref.~\onlinecite{2001PhyU...44..131K}, and also Ref.~\onlinecite{DeGottardi2011})  and the expression for $P_{\rm out}$ becomes less simple.}
\begin{equation}
P_{\text{out}}(\varphi)=i \Gamma_L \Gamma_R(\varphi),\label{eq:Pout}
\end{equation}
is locked to the value of $P_{\text{in}}$ in Eq.~(\ref{eq:parity}). 
In the thermodynamic limit, where each superconducting bank is sufficiently long, the MZM operators are well localized and commute with the lattice Hamiltonian~\eqref{eq:Hcf} even when the JJ is strongly interacting. Thus, for $N$ sufficiently large, the states $|j;\varphi\rangle$ are eigenstates of $P_{\rm out}(\varphi)$.

The computation of the expectation value  $P_{\text{out},i}\left(\varphi\right)\equiv\left\langle i;\varphi\left| P_{\text{out}}\right|i;\varphi\right\rangle $ returns Fig.~\ref{fig:TRB_spec}(c), showing that the parity is conserved and continuously defined for the blue and orange ground doublet across the full $4\pi$ evolution. The small kinks close to $\varphi=\pi, 3\pi$ are finite-size effects that should vanish for larger values of $N$.
Consequently, the crossings are protected and the $4\pi$ Josephson effect develops.

\begin{figure}[tb]
\includegraphics[width=1\columnwidth]{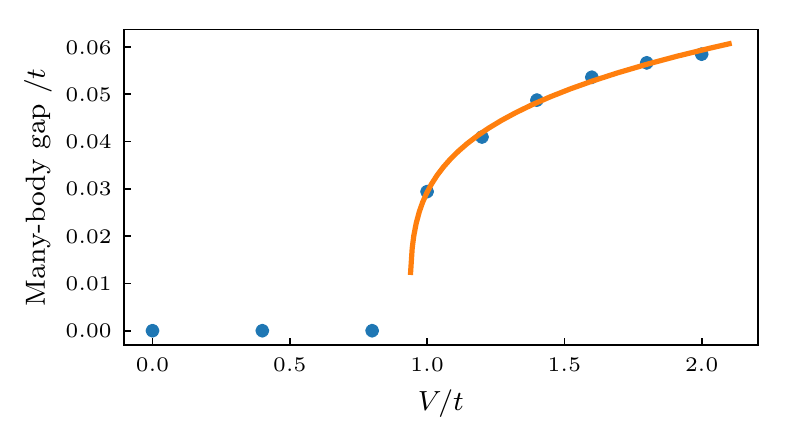}
\caption{Many-body gap of the interacting Kitaev chain JJ, calculated by exact diagonalization (blue disks). 
The parameter values are $\mu=0$, $\Delta_{0}=t/2$,
$N_{L}=8$ and $N=26$. The gap scales as $\sim (V-V_c)^{\gamma}$, where $V_c$ is the critical interaction strength for the gap opening. $V_c \simeq (0.94 \pm 0.02) t$, and $\gamma \simeq (0.25 \pm 0.03)$ are obtained from a power-law fit of the numerical data (orange solid curve).
\label{fig:MB_gap}}
\end{figure}

From the many-body spectrum and wavefunctions, we can obtain matrix elements of physical observables. 
With the cQED applications of the next section in mind, let us consider $N_{ij}\left(\varphi\right)\equiv\left\langle i;\varphi\left|\hat{N}\right|j;\varphi\right\rangle $, the matrix elements of the total number of particles $\hat{N}=\sum_{i}c_{i}^{\dagger}c_{i}$.
Figure ~\ref{fig:TRB_spec}(b) displays $N_{0j}$, where the state $i=0$ corresponds to the blue state in  Fig.~\ref{fig:TRB_spec}(a). Since the total number of particles is a sum over local operators, it cannot connect states with different values of the non-local operator $P_\text{out}$.
This is why $N_{01}\left(\varphi\right)$ and $N_{02}\left(\varphi\right)$ vanish.
\footnote{The computed values of $N_{01}\left(\varphi\right)$ are $\mathcal{O}\left(10^{-3}\right)$ for our parameters of choice. We performed a finite-size scale analysis of this quantity and found that it decreases in a oscillatory way, enveloped by a monotonically decaying function. This suggests that, in the thermodynamic limit (where there is no overlap between the well-localized inner and outer MZMs), $N_{01}\left(\varphi\right)$ vanishes,
with the local-fermion-parity a fully well-defined quantum number at low energies.} In contrast, $N_{03}$ is nonzero because the state $j=3$ has the same parity as $j=0$ and can thus be connected by a local and parity-preserving operator such as $\hat N$.\\

\subsection{Short-range Coulomb interactions\label{subsec:TR-Preserving-Interactions}}

The previous subsections have established the equivalence between the low-energy
properties of the QSH JJ and the Kitaev chain JJ at the non-interacting level. Here, we incorporate to the Hamiltonian (\ref{eq:Hcf}) the simplest possible interaction term, a first neighbor extended Hubbard interaction inside the normal region of the junction
\begin{equation}
H_{EH}=V\sum_{l=-N_L/2}^{N_L/2-2}n_{l}n_{l+1}.\label{eq:HEH}
\end{equation} 
Outside the junction, the proximity coupling to a three dimensional superconductor is assumed to screen away the interactions. 

In the continuum approximation,  the extended Hubbard interactions decompose into
\begin{widetext}
\begin{align}
\label{eq:dec}
\frac{c_{n}^{\dagger}c_{n}}{a} \frac{c_{n+1}^{\dagger}c_{n+1}}{a}
&\approx \rho\left(x\right)\rho\left(x+a\right)+\left[e^{i2k_{F}a}\left(\psi_{R}^{\dagger}\psi_{L}\right)\left(x\right)\left(\psi_{L}^{\dagger}\psi_{R}\right)\left(x+a\right)+h.c.\right]\nonumber \\
 & +\left[e^{-i2k_{F}\left(2x+a\right)}\left(\psi_{R}^{\dagger}\psi_{L}\right)\left(x\right)\left(\psi_{R}^{\dagger}\psi_{L}\right)\left(x+a\right)+h.c.\right]
\nonumber \\
& +\left[e^{-i2k_{F}x}\left(e^{-i2k_{F}a}\rho\left(x\right)\left(\psi_{R}^{\dagger}\psi_{L}\right)\left(x+a\right)+\left(\psi_{R}^{\dagger}\psi_{L}\right)\left(x\right)\rho\left(x+a\right)\right)+h.c.\right],
\end{align}
\end{widetext}
where $\rho\left(x\right)=:\psi_{R}^{\dagger}\psi_{R}:\left(x\right)+:\psi_{L}^{\dagger}\psi_{L}:\left(x\right)$
and the colons indicate normal ordering.
The terms in Eq.~(\ref{eq:dec}) coincide with the ones one would write for an interacting QSH edge with TR symmetry.
This establishes the equivalence between the QSH JJ and the Kitaev chain JJ at the interacting level. The first non-oscillating terms on the right hand side of Eq.~(\ref{eq:dec}) are known to renormalize the velocities of left- and right-moving fermions,~\cite{giamarchi} without opening spectral gaps. The second line of Eq.~(\ref{eq:dec}) (umklapp/pair-backscattering terms) and the third line (Friedel oscillating terms) are rapidly oscillating away from $\mu=0$ and $\mu=\pm2t$, respectively.
In perturbative renormalization group analyses, only the umklapp terms at half-filling ($\mu=0$) are seen to lead to a strong-coupling flow that indicates a gap opening in the low energy degrees of freedom.

Figure~\ref{fig:MB_spec}(a) displays the energy dispersion as a function of $\varphi$ for the 8 lowest-energy many-body states, obtained from  exact diagonalization. The blue, orange, green and red curves form the ground state multiplet, separated from the excited states (in grey) by a many-body gap. This gap develops at the 4-fold crossing of Fig.~\ref{fig:MB_no_int} and scales as $\sim (V-V_c)^{\gamma}$, with $V_c/t \simeq 0.94$  and $\gamma\simeq0.25$ obtained by a power-law fit of the numerical data (see Fig.~\ref{fig:MB_gap}). 

We remark that $V/t>1$ is a strong interaction regime, likely hard to achieve in real systems. Also, even in this strong coupling regime, the many-body gap is a small fraction of the bandwidth. We have checked numerically that the many-body gap is not reduced when $\mu\neq 0$. At first sight, this finding is surprising from the point of view of a perturbative analysis. 
One possible explanation is that the gap may be originating from oscillatory umklapp or Friedel terms, because these oscillations are not sufficiently fast to average out in our weak links of mesoscopic size.  Another possible explanation is that the perturbative arguments arguing for the irrelevance of the oscillatory Friedel and umklapp terms break down in the strong coupling regime, where the interaction strength exceeds the bandwidth. 

The colors in Fig.~\ref{fig:MB_spec}(a) are in one-to-one correspondence with the mean values of $P_{\text{out}}$ in Figs.~\ref{fig:MB_spec}(c) and (d).
This parity conservation protects the band crossings at $\varphi=\pi$ mod $2\pi$, while the crossings at $\varphi=0$ mod $2\pi$ are preserved by TR symmetry. Thus, the ground state multiplet is $8\pi$-periodic and so is the Josephson effect. In anticipation to the next section, let us once again consider the off-diagonal elements of the total number operator $\hat{N}$ for the states belonging to the ground state multiplet, as displayed in Fig.~\ref{fig:MB_spec}(b). The fact that $N_{01}(\varphi)\simeq N_{02}(\varphi)\simeq 0$ for all values of $\varphi$ can be attributed to the conservation of the local fermion parity. Indeed, the extended Hubbard interactions, which act again locally and only inside the weak-link, commute with $P_{\text{out}}$. In contrast, $N_{03}\left(\varphi\right)$ is finite for all values of $\varphi$ away from $\varphi=2\pi\,\mod4\pi$ because the states $j=0$ and $j=3$ carry the same parity. The crossings between $j=0$ and $j=3$ are, however, still protected by the effective low-energy TR symmetry $\mathcal{T}_-$, and that translates into the vanishing of $N_{03}\left(\varphi\right)$ at $\varphi=2\pi, 6\pi$. 

\section{Fractional Josephson effects in cQED architectures\label{sec:Circuit-QED}}

\begin{figure}[tb]
\includegraphics[width=0.75\columnwidth]{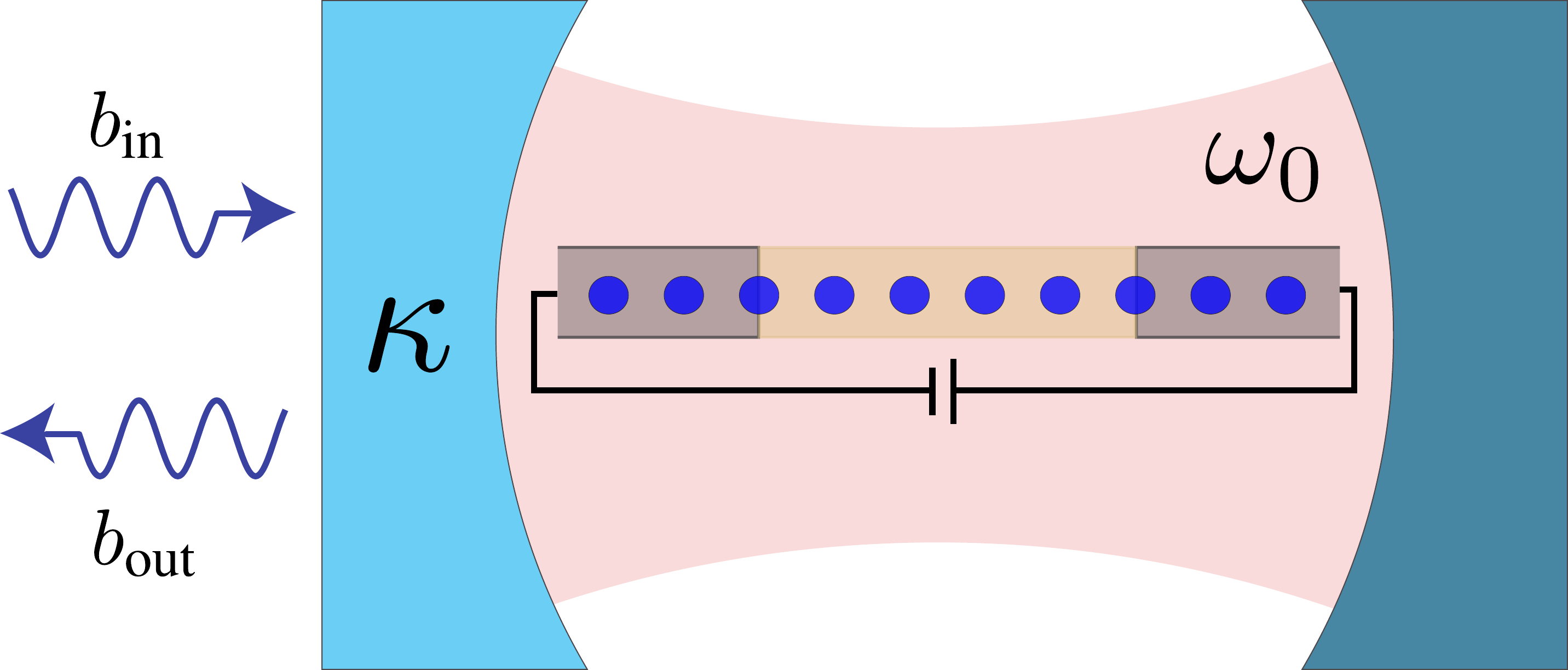}
\caption{Cartoon of a cQED architecture. 
A cavity resonator of frequency $\omega_0$ and linewidth $\kappa_0$ contains an interacting topological JJ.
The cavity is partially transmitting on a single side, so that it can be probed by input/output fields.
\label{fig:cQED-architecture-schematics}}
\end{figure}

Due to their high sensitivity, flexibility and non-invasive probing, cQED platforms have been proposed for the study and detection of topological phases in Josephson junctions.\cite{Cottet:2013ai,PhysRevB.92.245432,PhysRevB.94.115423} The general approach of cQED, as illustrated in Fig.~\ref{fig:cQED-architecture-schematics}, consists of two steps: (i) the placement of the circuit one wishes to study inside a cavity resonator and (ii) the measurement of reflectances and transmittances between in-/out-signal microwave modes inserted in the cavity through a waveguide. The in-/out-modes couple to the cavity photons, whose dynamics is in turn influenced by the dynamics of the circuit of interest.

In this section, we present an exact diagonalization calculation of certain cQED observables in a topological JJ.
Our study goes beyond earlier theoretical works by incorporating strong short-range Coulomb interactions, crucial for the emergence of the $8\pi$-periodic Josephson effect.

\subsection{Input/output formalism \label{subsec:input-output}}
We consider a Hamiltonian with three components: probing fields, a cavity and a topological Josephson junction, 
\begin{align}
H & =H_{S}+H_{I}+\omega_{0}a^{\dagger}a\nonumber \\
 & +\sum_{n}\Omega_{n}b_{n}^{\dagger}b_{n}-i\sum_{n}\lambda_{n}\left(a^{\dagger}b_{n}-ab_{n}^{\dagger}\right).
\end{align}
Here, $a^{(\dagger)}$ and $b_{n}^{(\dagger)}$ are the annihilation (creation) 
operators for cavity photons and the mode $n$ of the probing field, respectively; $\Omega_{n}$ and
$\lambda_{n}$ are the frequencies of the probe fields and the cavity-probe coupling
constants, respectively; $\omega_{0}$ is the resonance frequency of the empty cavity. The Hamiltonian $H_{S}$ describes the JJ,
\begin{align}
H_{S} & =H_{JJ}\left(\varphi\right)+H_{\alpha},
\end{align}
where $\alpha=SB$ (cf. Eq.~(\ref{eq:HSIS})) or $EH$ (cf. Eq.~(\ref{eq:HEH})), depending on whether we are dealing with the $\mathbb{Z}_{2}$
or $\mathbb{Z}_{4}$ fractional Josephson effect.
Also, we consider a capacitive coupling between the junction
and cavity, 
\begin{equation}
H_{I}=g\hat{N}\left(a+a^{\dagger}\right),
\end{equation}
where $g$ is a coupling constant, and $\hat{N}=\sum_{i}c_{i}^{\dagger}c_{i}$
is the total number operator.

The dynamics of the cavity fields can be obtained
by the standard input/output formalism,~\cite{MilburnBook} yielding 
\begin{equation}
\dot{\tilde{a}}\left(t\right)=-i\left[\tilde{a}\left(t\right),\tilde{H}_{I}\right]-\left(i\omega_{0}+\frac{\kappa_0}{2}\right)\tilde{a}\left(t\right)+\sqrt{\kappa_{0}}\tilde{b}_{\rm in}\left(t\right).\label{eq:inputoutput}
\end{equation}
Here, tildes denote operators written in the Heisenberg picture,
$\kappa_{0}$
is the cavity damping constant due to the coupling with the probe and 
\begin{equation}
\label{eq:bin}
\tilde{b}_{\rm in}\left(t\right)\equiv\sum_{n}\lambda_n\tilde{b}_{n}\left(t_{0}\right)e^{-i\Omega_{n}\left(t-t_{0}\right)}
\end{equation}
is the input field with $t_0$ a reference time.
In Eq.~(\ref{eq:bin}), the sum over the modes is constrained to $\Omega_{n}\approx\omega_{0}$.
The input field is related to the output field $b_{\rm out}$  by the boundary condition 
\begin{equation}
\tilde{b}_{\rm in}\left(t\right)+\tilde{b}_{\rm out}\left(t\right)=\sqrt{\kappa_{0}}\tilde{a}\left(t\right).\label{eq:bcs}
\end{equation}
When $b_{\rm in}$ is used to drive the
cavity, the readout of $b_{\rm out}$ enables to measure 
the cavity frequency and linewidth. The commutator in Eq.~(\ref{eq:inputoutput}) forecasts that the dynamics of the junction will be intertwined with that of the cavity. To second order in $g$ and in the rotating-wave approximation, Eq.~(\ref{eq:inputoutput}) becomes (cf. Appendix~\ref{sec:Cavity-QED-derivations})
\begin{equation}
\dot{\tilde{a}}\left(t\right)=-\left(i\omega_R+\frac{\kappa_R}{2}\right)\tilde{a}\left(t\right)+\sqrt{\kappa_{0}}\tilde{b}_{\rm in}\left(t\right)
-ig N^{i}\left(t\right),\label{eq:cav_eff}
\end{equation}
where $\omega_R=\omega_0+\bar{\omega}(\varphi)$ and $\kappa_R=\kappa_0+\bar{\kappa}(\varphi)$ are the renormalized cavity frequency and linewidth, whereas 
$N^{i}\left(t\right)$ is the number operator in the interaction
picture. Thus, the junction induces a $\varphi$-dependent pull $\bar{\omega}\left(\varphi\right)$
in the resonance frequency of the cavity, in addition to a $\varphi$-dependent
change $\bar{\kappa}(\varphi)$ in the cavity linewidth.~\cite{PhysRevB.94.115423}
The explicit expressions for $\bar{\omega}(\varphi)$ and $\bar{\kappa}(\varphi)$ are shown below.
The last term in Eq.~(\ref{eq:cav_eff}) is an extra driving term for the cavity which, as shown in Appendix~\ref{sec:Cavity-QED-derivations}, may be ignored because it contributes only at zero frequency. 

The fractional Josephson effects manifest themselves in the $\varphi$-dependence of $\omega_R$ and $\kappa_R$.
If the rate at which $\varphi$ is varied is faster than all the energy relaxation rates of the quasiparticles but smaller than the topological energy gap, 
$\omega$ and $\kappa$ are $4\pi$- ($8\pi$-) periodic functions of $\varphi$ when the junction hosts a $\mathbb{Z}_2$ ($\mathbb{Z}_4$) Josephson effect. One way to realize this condition is through the application of an appropriate dc voltage bias $V$ across the junction, under which $\varphi=\varphi_0+2 e V t/\hbar$ evolves with time $t$. In the remainder of this section, we compute $\bar{\omega}(\varphi)$ and $\bar{\kappa}(\varphi)$ and propose an experiment to capture their anomalous periodicities via the input/output fields.

\subsection{Cavity frequency pull \label{subsec:pull}}

At zero temperature, the expression for the cavity frequency pull reads (cf. Appendix~\ref{sec:Cavity-QED-derivations})
\begin{equation}
\bar{\omega}\left(\varphi\right)=2g^{2}\sum_{j\neq0}\left|N_{0j}\left(\varphi\right)\right|^{2}\frac{\Delta E_{0j}\left(\varphi\right)}{\left(\Delta E_{0j}\left(\varphi\right)\right)^{2}-\omega_{0}^{2}},\label{eq:freq_pull}
\end{equation}
where $\Delta E_{0j}\left(\varphi\right)= E_{0}\left(\varphi\right)-E_{j}\left(\varphi\right)$,
$N_{0j}\left(\varphi\right)=\left\langle 0; \varphi\left|\hat{N}\right|j;\varphi\right\rangle $, and  $\left|0;\varphi\right\rangle $ is the many-body state whose energy is the lowest of all when $\varphi=0$ (the blue band in either Fig.~\ref{fig:TRB_spec}a or Fig.~\ref{fig:MB_spec}a). Replacing the state $\left|0;\varphi\right\rangle $ by any other states in the ground state multiplet amounts
to an inconsequential shift of $\varphi$ by a multiple of $2\pi$ in Eq.~(\ref{eq:freq_pull}).
The sum in $j$ is over all other states, scattering states included.
In our numerical calculations, we truncate the sum to the 8 lowest-energy many-body states.

\begin{figure}[tb]
\begin{centering}
\includegraphics[width=1\columnwidth]{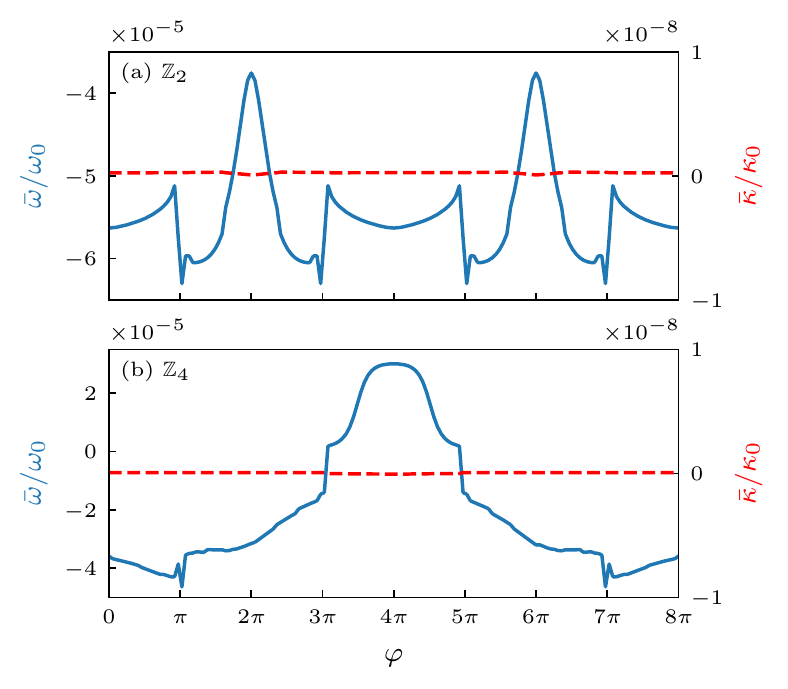}
\par\end{centering}
\caption{Renormalization of the cavity resonance frequency (blue, full lines) and
linewidth (red, dashed lines), calculated by exact diagonalization of the Kitaev chain JJ. 
(a) Non-interacting JJ with broken time-reversal
symmetry. (b) Interacting JJ with time-reversal symmetry.
The bare cavity resonance frequency is chosen as $\omega_{0}=4\times10^{-3}\Delta_{c}$, parametrically smaller than the energy gaps separating the ground state multiplets from the excited states.
The broadening of the delta functions in Eq.~(\ref{eq:kappa_ren}) is taken to be of the order of the bare cavity linewidth $\kappa_0\simeq 10^{-3} \omega_0$.
For the cavity-junction coupling strength, we use $g=\omega_{0}/10$.~\cite{Bosman} The periodicity of the cavity pull in the superconducting phase difference follows that of many-body wavefunctions of the problem. When $\omega_{0}$ is large compared to the disorder broadening of the ABS, but smaller than the energy gaps separating the ground state multiplet from the excited states, the conservation of the local-fermion-parity (as well as time-reversal symmetry, in the case of the $\mathbb{Z}_4$ Josephson effect) results in a negligible renormalization of the cavity linewidth.
\label{fig:cavipull}}
\end{figure}

Figure~\ref{fig:cavipull} displays $\bar{\omega}(\varphi)$ for JJs hosting $\mathbb{Z}_{2}$ and $\mathbb{Z}_4$ Josephson effects. The cavity frequency pull is either $4\pi$- or $8\pi$-periodic in $\varphi$. For the $8\pi$ scenario, $\bar{\omega}$ can actually become positive, in contrast
with the $4\pi$ periodic case. 
The origin of this difference comes from transition matrix elements between states of the same local fermion parity. In JJs hosting the $\mathbb{Z}_4$ Josephson effect, the fact that $N_{03}(\varphi)\neq 0$ for generic $\varphi$ and $\Delta E_{03}(\varphi)>0$ for certain intervals of $\varphi$ (cf. Fig.~\ref{fig:MB_spec}) is responsible for $\bar{\omega}(\varphi)>0$ in those intervals. This situation is not realized in JJs that host the $\mathbb{Z}_{2}$ Josephson effect, where the conservation of the local fermion parity forbids transitions with $\Delta E_{0j}>0$.

\subsection{Cavity linewidth renormalization \label{subsec:linewidth}}

At zero temperature, the renormalization
of the cavity linewidth is given by
\begin{align}
\bar{\kappa}(\varphi) & =4\pi g^{2}\sum_{j\neq0}\left|N_{0j}\left(\varphi\right)\right|^{2}\label{eq:kappa_ren}\\
 & \times\left[\delta\left(\Delta E_{0j}\left(\varphi\right)+\omega_{0}\right)-\delta\left(\Delta E_{0j}\left(\varphi\right)-\omega_{0}\right)\right].\nonumber 
\end{align}
The Dirac deltas are to be broadened into Lorentzians by effects such
as disorder and feedback of the cavity dynamics into the junction
energies, which will be considered here only phenomenologically. 
\footnote{More sophisticated treatments of disorder have been employed recently, but only for noninteracting junctions; see e.g. M. Trif, O. Dmytruk, H. Bouchiat, R. Aguado and P. Simon, Phys. Rev. B {\bf 97}, 041415(R) (2018).}

The values of $\bar{\kappa}$ calculated by exact diagonalization are presented in 
Fig.~\ref{fig:cavipull} (red dashed lines). Importantly, when $\omega_{0}$ is large compared to the disorder broadening of the ABS but smaller than the TR-breaking gap from $\delta\mu$ or the TR-preserving gap
due to interactions (a circumstance believed to be realistic), we find $\bar{\kappa}\simeq 0$. This null result has a simple explanation. Because $\omega_{0}$ is small compared to typical ABS energy scales, the Dirac delta functions in Eq.~(\ref{eq:kappa_ren}) are satisfied only very close
to the crossings. But the matrix elements of $N_{0j}\left(\varphi\right)$ are vanishingly small at the crossings, due to the conservation of either the local-fermion-parity or time-reversal.

\subsection{Phase shift\label{subsec:shift}}

Having found how the topological JJ influences key physical properties of the cavity, we now focus on how to access these. The Fourier transform of Eq.~(\ref{eq:cav_eff}) and the boundary conditions in Eq.~(\ref{eq:bcs}) allow to relate the outgoing signal and incoming signals in a single-sided partially transparent cavity. Neglecting zero-frequency contributions, we obtain
\begin{equation}
\label{eq:outin}
\left\langle \tilde{b}_{\rm out}\left(\omega\right)\right\rangle =re^{i\alpha}\left\langle \tilde{b}_{\rm in}\left(\omega\right)\right\rangle ,
\end{equation}
where $r$ is the reflection coefficient obeying 
\begin{equation}
r^{2}=\frac{\left(\omega-\omega_{R}\right)^{2}+\left(\frac{\kappa_{0}-\bar{\kappa}}{2}\right)^{2}}{\left(\omega-\omega_{R}\right)^{2}+\left(\frac{\kappa_{R}}{2}\right)^{2}}
\end{equation}
and
\begin{equation}
\alpha=\arg\left[(\kappa_{0}^{2}-\bar{\kappa}^{2})/4-(\omega-\omega_{R})^{2}+i\kappa_{0}(\omega-\omega_{R})\right]
\end{equation}
is the phase shift.
The quantities $r$ and $\alpha$ are directly measurable in experiments.
As $\bar{\kappa}=0$ (cf. preceding subsection), $r^{2}=1$ for all values of $\varphi$. Hereafter, we concentrate on $\alpha$.

\begin{figure}[tb]
\includegraphics[width=1\columnwidth]{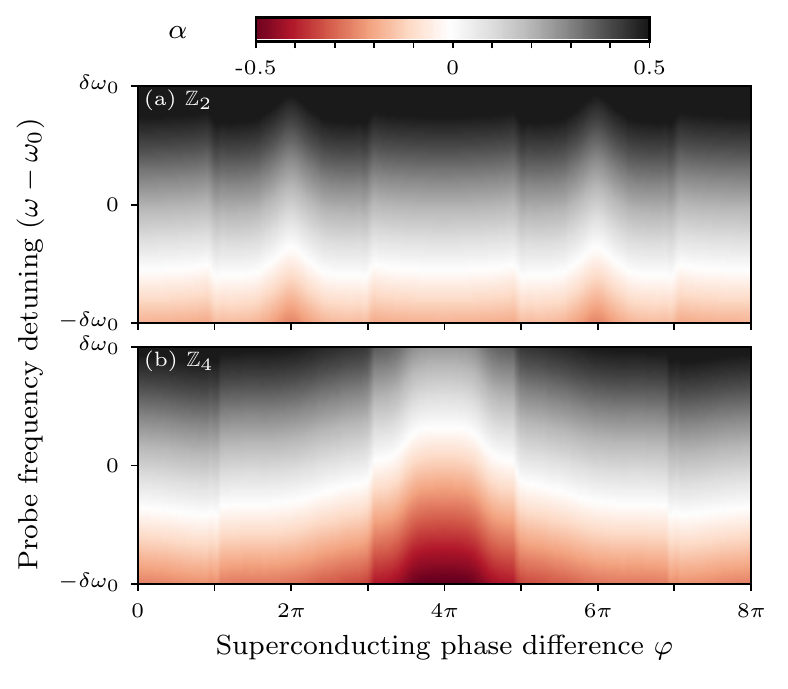}

\caption{Phase-shift $\alpha(\omega)$ between input and output signals for a single-sided cavity
containing a Kitaev chain JJ. 
We show $\alpha$ for $\omega\in(\omega_0-\delta\omega_0, \omega_0+\delta\omega_0)$, where $\omega_0$ is the resonance frequency of the empty cavity and $\delta\omega_{0}=10^{-1}\kappa_{0}$.
Top panel: non-interacting JJ with broken time-reversal symmetry. 
Bottom panel: interacting JJ with time-reversal symmetry.
\label{fig:phaseshift}}
\end{figure}

Figure~\ref{fig:phaseshift} shows the behaviour of $\alpha$ as function of $\varphi$ at frequencies around $\omega_{0}$. 
The phase shift changes sign as the frequency of the probe crosses the resonance frequency. This enables a precise determination of $\omega_R$ and its anomalous periodicity through reflectrometry measurements.

One approach to carry out the experimental verification of Fig.~\ref{fig:phaseshift} consists of the following steps: (1) Measure $\alpha$ as a function of the probing frequency $\omega$ in the absence of current and bias voltages. The value of $\omega$ at which $\alpha$ changes sign constitutes $\omega_R$ at $\varphi=0$. (2) Choose a window of frequency $\delta\omega_0$ around the $\varphi=0$ value of $\omega_R$. For each value of frequency inside this window, let $\varphi$ evolve in time while continuosuly measuring $\alpha$. The phase evolution is best accomplished by a dc voltage bias $V$, under which $\dot{\varphi}=2e V/\hbar$ is constant. This has the advantage of knowing how much $\varphi$ has wound in a given measurement time, thereby allowing to extract the periodicity of $\alpha$ in $\varphi$.

In order to observe the anomalous periodicities, the measurement time in step (2) must be shorter than the energy relaxation rate, longer than $4\pi\hbar/(e V)$ (so that the phase winds at least by $8\pi$), and  much longer than the inverse of the data acquisition rate (which is $\simeq 1$ ns in state of the art experiments~\cite{Walter}).
In addition, $2 e V$ must be smaller than (i) the energy gap separating the ground state multiplet from the excited states (to avoid Landau-Zener tunneling away from the ground state), and (ii) $\hbar\omega_0$ (to prevent that the inelastic tunneling of Cooper pairs generates photons at the cavity's frequency). All of these conditions are simultaneously satisfiable in view of recent reports~\cite{2017arXiv171101645H} of long ($\simeq 0.1$ ms) energy relaxation times in Al-coated InAs nanowires.

\section{Summary and conclusions\label{sec:Final-Remarks}}

We have presented an exact diagonalization study of fractional Josephson effects in interacting topological Josephson junctions (JJs). By a careful comparison with a continuum low-energy version of the problem, we have established that JJs created out of Kitaev chains can be used to simulate JJs created at the edges of quantum spin-Hall insulators. Central to this equivalence is an emergent time-reversal symmetry squaring to $-1$ in the low-energy description of
the lattice problem. The existence of this effective symmetry is contingent on having a perfect transparency in the Kitaev chain JJ.

The use of Kitaev chain JJs to simulate quantum spin-Hall JJs offers two advantages.
First, it extends the $8\pi$-periodic Josephson effect to systems other than quantum spin Hall insulators, where it was originally proposed. In this regard, the ongoing advances towards the engineering of Kitaev chains,~\cite{Hao_natcomms,Chang_natcomms,Lutchyn_NatRev} together with  gate-tuned transparencies of up to 98\% reported in Al-coated InAs nanowires, \cite{2017arXiv171101645H} presage the realization of Kitaev chain JJs of high transparency.

Another advantage of our lattice simulations is that they give access to physical observables that are difficult to compute using continuum analytical approaches from earlier works. To illustrate this point, we have considered an interacting Josephson junction coupled to a microwave resonator and have calculated the renormalizations of the cavity's resonance frequency and linewidth.
We have found that the cavity linewidth is approximately unchanged by the presence of the junction for a reasonable range of physical parameters, while the cavity frequency displays $4\pi$- and $8\pi$-periodic features that may be accessed by measuring the phase shift between incoming and outgoing signals.

For future work, it will be interesting to investigate signatures of the $\mathbb{Z}_2$ and $\mathbb{Z}_4$ Josephson effects in higher-order photon correlation functions.

\begin{acknowledgments}
We thank D. S\'{e}n\'{e}chal, V. L. Quito , J. Teo, B. Dou\c{c}ot, P. Ghaemi,
R. Pereira, J.O. Simoneau, B. Reulet, and F. Zhang for insightful discussions
and suggestions. This research has been financed by the Canada First Research Excellence Fund, the Natural Science and Engineering Council of Canada, and the Fonds de Recherche du Qu\'ebec Nature et Technologies. 
Numerical calculations were done with computer resources from Calcul Qu\'ebec and Compute Canada. 
\end{acknowledgments}

\appendix
\begin{figure*}[t]
\begin{centering}
\includegraphics[width=0.33\textwidth]{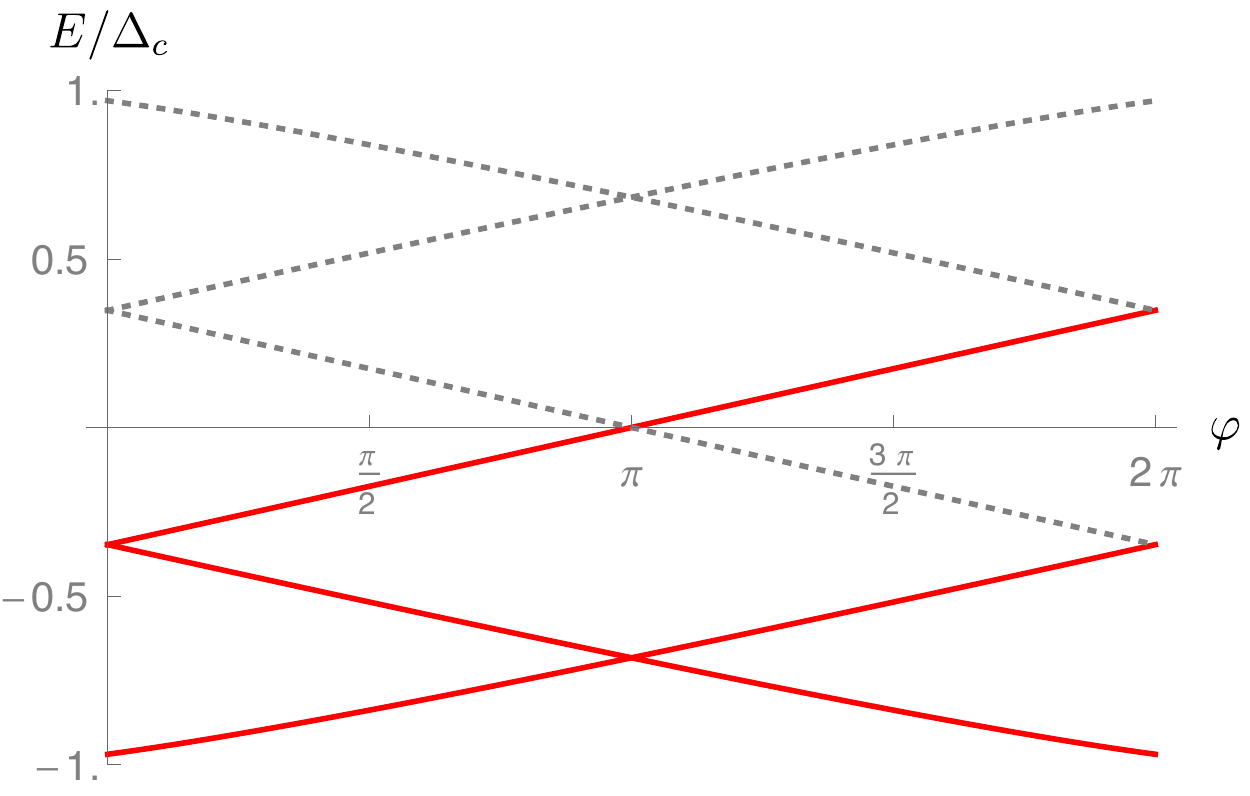}\includegraphics[width=0.33\textwidth]{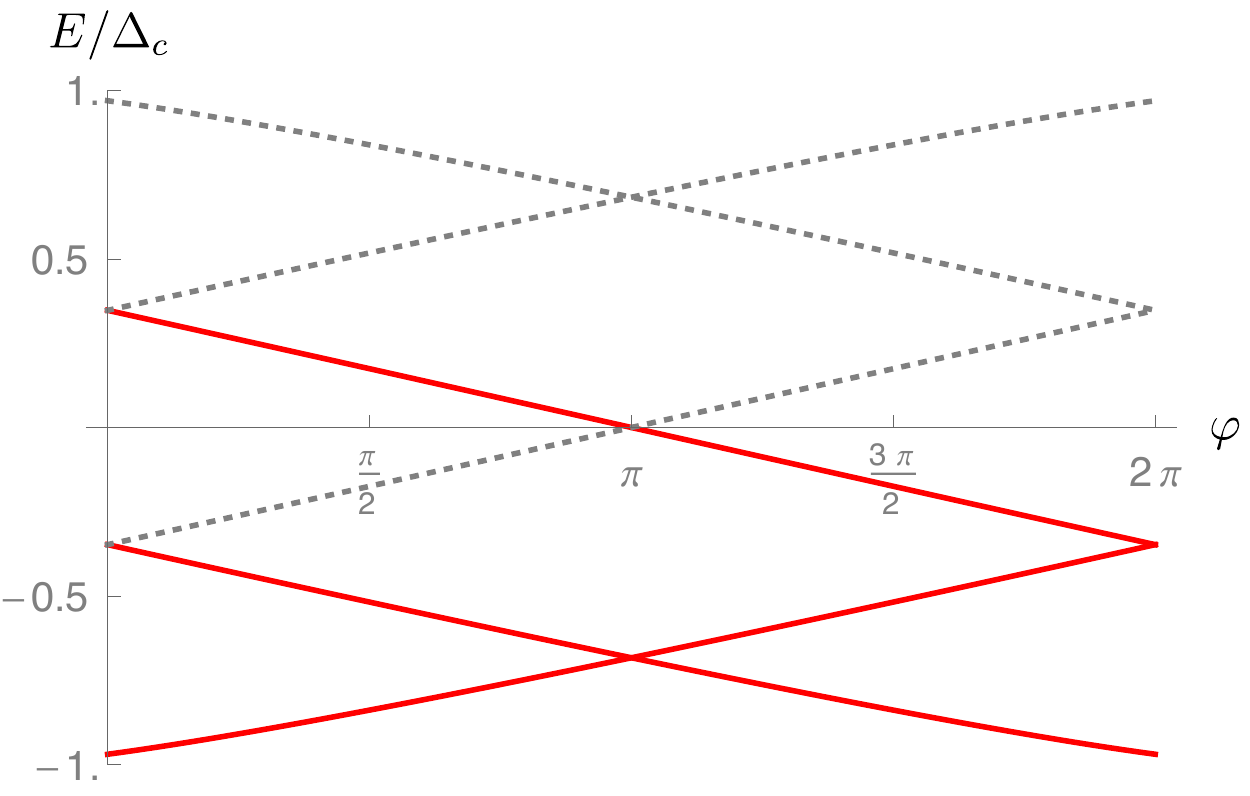}\includegraphics[width=0.33\textwidth]{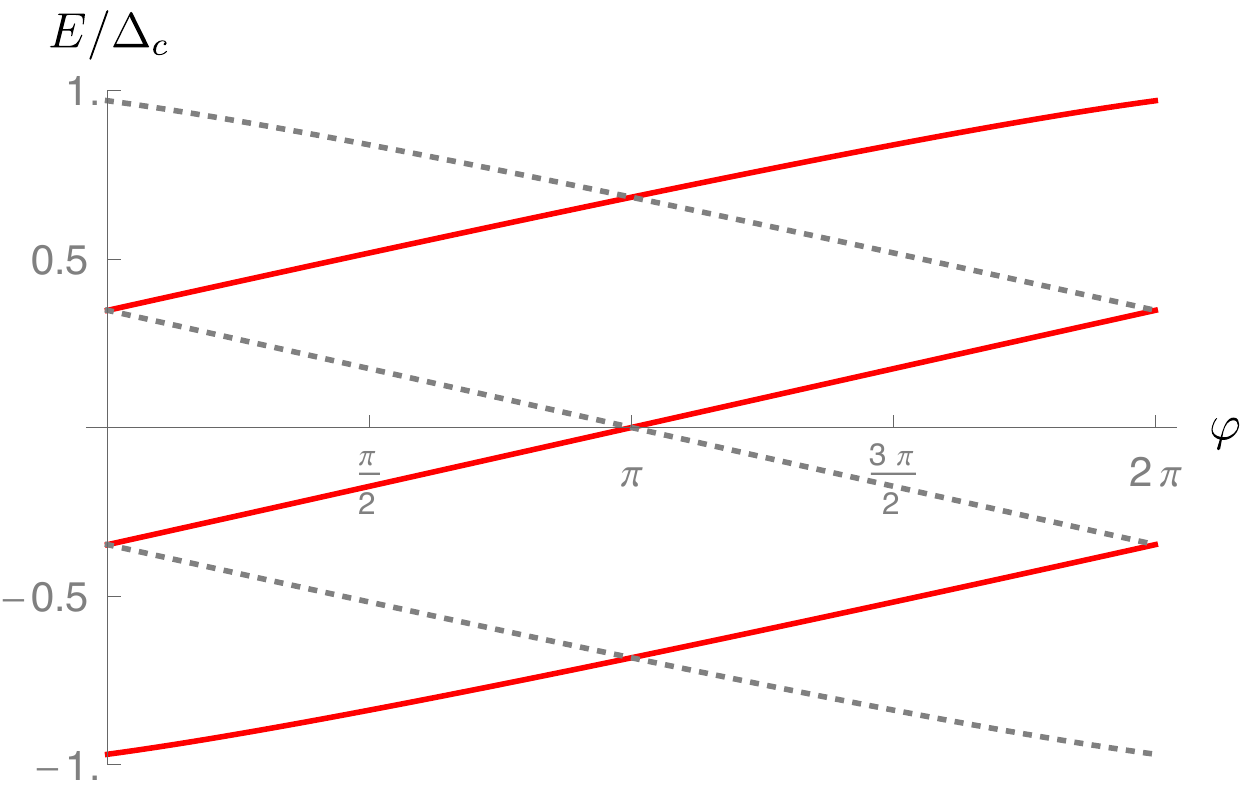}
\par\end{centering}
\begin{centering}
\includegraphics[width=0.33\textwidth]{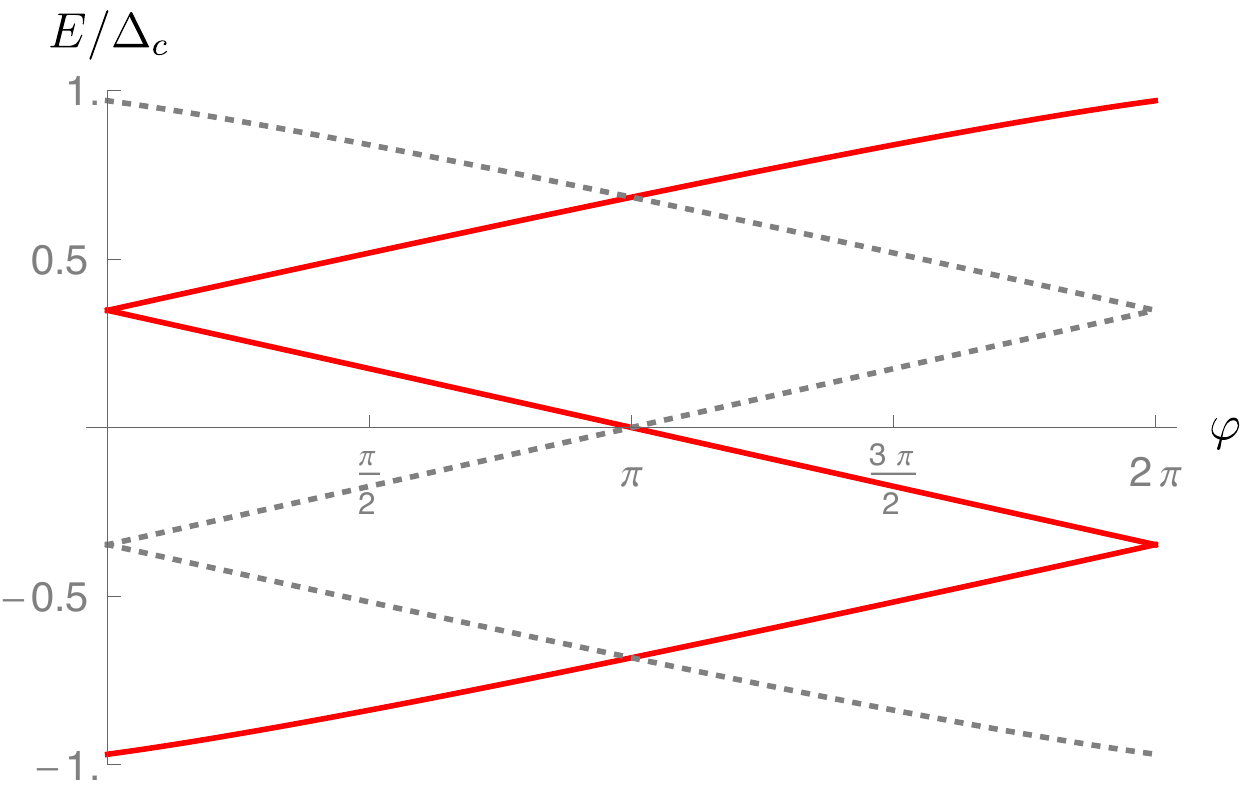}\includegraphics[width=0.33\textwidth]{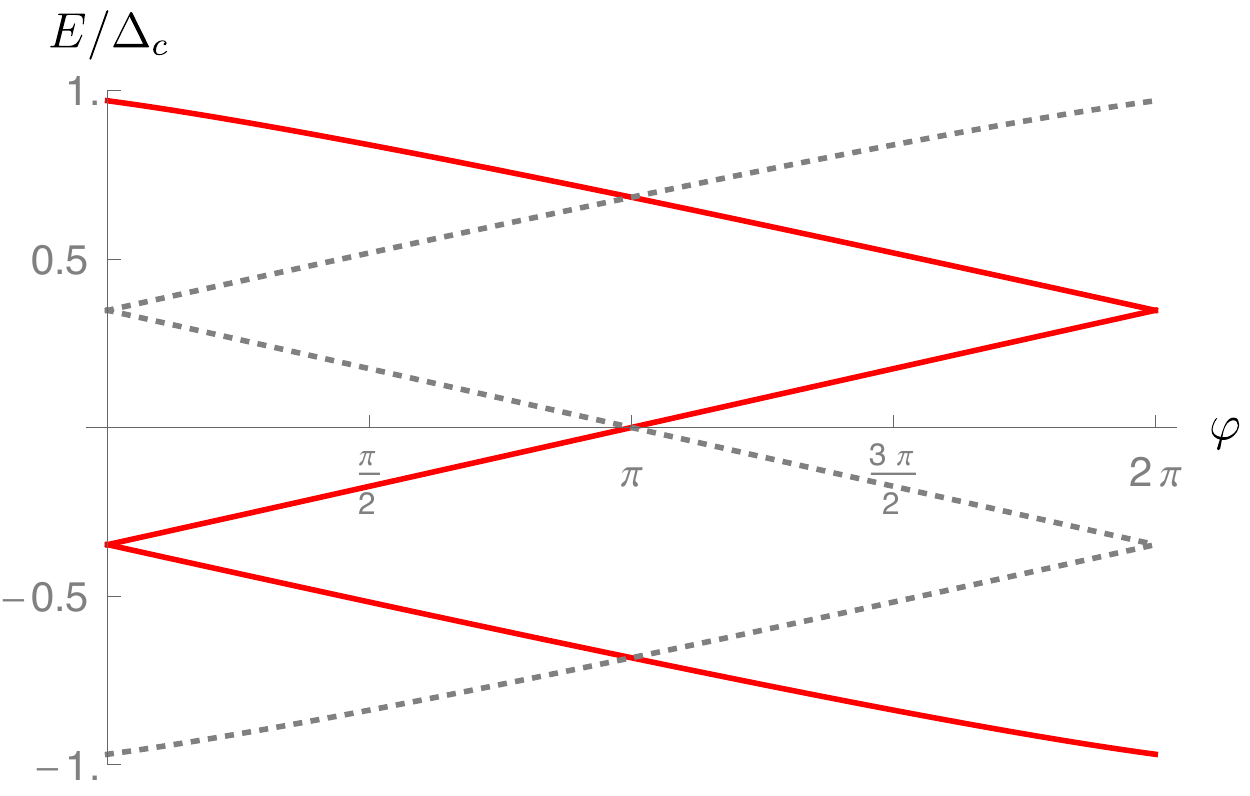}\includegraphics[width=0.33\textwidth]{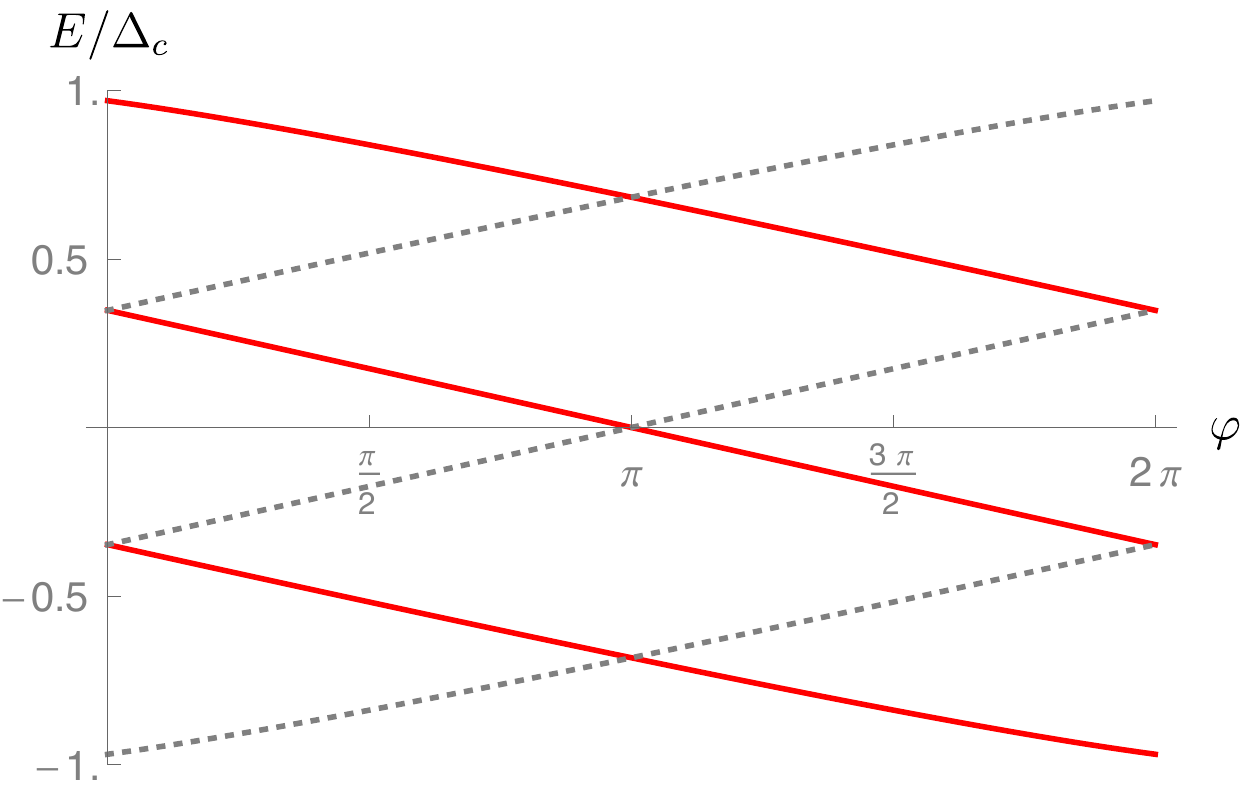}
\par\end{centering}
\caption{Pictorial representation of the non-interacting many-body states in Eq.~(\ref{eq:mbstates}).
The first row displays states $\left|0;\varphi\right\rangle ,\,\left|1;\varphi\right\rangle ,\,\left|2;\varphi\right\rangle $
and second row displays states $\left|3;\varphi\right\rangle ,\,\left|4;\varphi\right\rangle ,\,\left|5;\varphi\right\rangle $,
from left to right in both cases.
The dashed grey lines (red full lines) correspond to empty (occupied) single-particle Andreev bound states. Scattering states are not shown.
 \label{fig:MB_non_int_states}}
\end{figure*}
\section{Single-particle many-body states\label{subsec:Single-particle-many-body-states}}

The knowledge of the single-particle Andreev-bound-states allows for an explicit construction of non-interacting low-energy many-body
states. 
Without applying the Nambu constraint, these many body states  are obtained by the introduction of positive energy particles
and destruction of their corresponding negative energy particle-hole
symmetric partners.
The first six states read 
\begin{align}
\left|0;\varphi\right\rangle  & =\left[\prod_{n<0}a_{n,+}^{\dagger}\right]\left[\prod_{n<0}a_{n,-}^{\dagger}\right]\left[a_{0,-}^{\dagger}\right]\left|\Omega_{e}\right\rangle \nonumber \\
\left|1;\varphi\right\rangle  & =a_{0,+}^{\dagger}a_{0,-}\left|0;\varphi\right\rangle \nonumber \\
\left|2;\varphi\right\rangle  & =a_{1,-}^{\dagger}a_{-1,+}\left|0;\varphi\right\rangle \nonumber \\
\left|3;\varphi\right\rangle  & =a_{1,-}^{\dagger}a_{0,+}^{\dagger}a_{-1,+}a_{0,-}\left|0;\varphi\right\rangle \nonumber \\
\left|4;\varphi\right\rangle  & =a_{1,+}^{\dagger}a_{-1,-}\left|0;\varphi\right\rangle \nonumber \\
\left|5;\varphi\right\rangle  & =a_{1,+}^{\dagger}a_{0,+}^{\dagger}a_{-1,-}a_{0,-}\left|0;\varphi\right\rangle ,\label{eq:mbstates}
\end{align}
where $\left|\Omega_{e}\right\rangle $ is the non-superconducting
electron Fermi sea and the $\varphi$ dependence of the operators
has been omitted. These states are pictorially represented in Fig.~\ref{fig:MB_non_int_states}. 

To incorporate the Nambu constraint, we have to define a reference set
of operators and enforce the particle-hole operation
\begin{align}
{\cal C}\Psi\left(x\right){\cal C}^{-1} & =\rho_{y}\tau_{y}\left[\Psi^{\dagger}\left(x\right)\right]^{T}=\Psi\left(x\right).\label{eq:NambuConstr}\\
\end{align}
The excitations can then be constructed as 
\begin{align}
\left|0;\varphi\right\rangle  & =\left[\prod_{n>0}a_{n,+}\right]\left[\prod_{n>0}a_{n,-}\right]\left[a_{0,+}\right]\left|\Omega_{e}\right\rangle \nonumber \\
\left|1;\varphi\right\rangle  & =a_{0,+}^{\dagger}\left|0;\varphi\right\rangle \nonumber \\
\left|2;\varphi\right\rangle  & =a_{1,-}^{\dagger}\left|0;\varphi\right\rangle \nonumber \\
\left|3;\varphi\right\rangle  & =a_{1,-}^{\dagger}a_{0,+}^{\dagger}\left|0;\varphi\right\rangle \nonumber \\
\left|4;\varphi\right\rangle  & =a_{1,+}^{\dagger}\left|0;\varphi\right\rangle \nonumber \\
\left|5;\varphi\right\rangle  & =a_{1,+}^{\dagger}a_{0,+}^{\dagger}\left|0;\varphi\right\rangle .
\end{align}

Enforcing the Nambu constraint, the normal ordered Hamiltonian for
the junction reads
\begin{align}
:H_{JJ}\left(\varphi\right): & =E_{0,+}\left(\varphi\right)a_{0,+}^{\dagger}a_{0,+}+\sum_{n>0,\tau}E_{n,\tau}\left(\varphi\right)a_{n,\tau}^{\dagger}a_{n,\tau}\nonumber \\
 & -\frac{1}{2}\left[E_{0,+}\left(\varphi\right)+\sum_{n>0,\tau}E_{n,\tau}\left(\varphi\right)\right]\nonumber \\
 & +\frac{1}{2}\left[E_{0,+}\left(0\right)+\sum_{n>0,\tau}E_{n,\tau}\left(0\right)\right],
\end{align}
which means that one first has to chose a reference state (here $\left|0;\varphi=0\right\rangle $),
normal order with respect to it, and then consider the evolution of
the phase $\varphi$ to other values. 

\begin{figure}[tb]
\begin{centering}
\includegraphics[width=0.8\columnwidth]{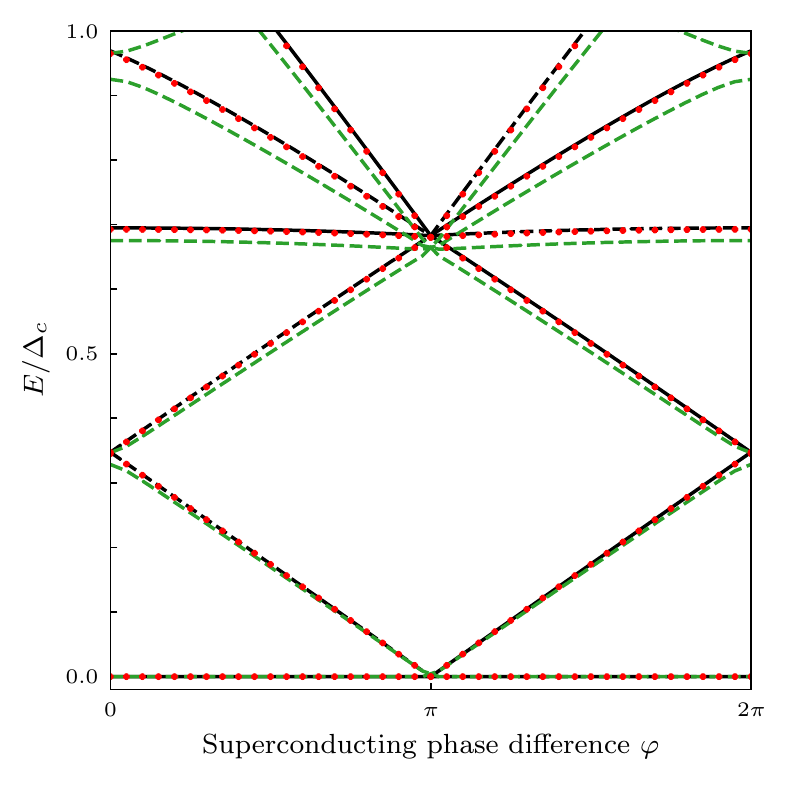}
\par\end{centering}
\caption{Comparison between the analytical and exact diagonalization results for the non-interacting
many-body spectrum.
Only energy differences with respect to the lowest-energy states are considered at each $\varphi$. 
Black solid lines come from the continuum theory, red dots from the single-body lattice
diagonalization and green dashes from the many-body exact diagonalization;
parameters are the same as used in the main text. The absence of the
contributions from the continuum of scattering states leads to a better
matching between both approaches, in comparison with Fig.~\ref{fig:MB_no_int}.
Splittings at $\varphi=0,\pi$ due to $\mathcal{T}_-$-breaking terms are reduced for the single-particle calculation by considering scaled parameters such that $\xi_{0}/a$ is scaled by a 
a factor of $3$.
\label{fig:MB_latt_vs_EFT}}
\end{figure}

Figure~\ref{fig:MB_latt_vs_EFT} compares the results from the lattice and continuum models for the low-energy many-body spectrum. For each value of $\varphi$, we plot the excitation energies with respect to the ground state. For the continuum model, this is equivalent to normal ordering at each value of $\varphi$ separately. This has the merit of cancelling out the contribution from the scattering states.
Consequently, the agreement between the lattice and continuum models is better than in Fig.~\ref{fig:MB_no_int}. It is also worth noting that Fig.~\ref{fig:MB_latt_vs_EFT} corresponds to the energy peaks in the tunneling density of states of the junction,\cite{PhysRevLett.113.036401} up to a selection rule that bars transitions to excited states with the same total fermion parity as the ground-state.

\section{Cavity renormalization and input-output calculation\label{sec:Cavity-QED-derivations}}

In this Appendix, we show the derivation leading to Eqs.~(\ref{eq:cav_eff}) and (\ref{eq:outin}). The starting point is the Hamiltonian
\begin{align}
H & =H_{S}\left(\varphi\right)+H_{\text{cav}}+H_{I},
\end{align}
where $H_S$ is the JJ Hamiltonian with many-body eingenstates $|j;\varphi\rangle$ and many-body eigenvalues $E_j(\varphi)$, 
\begin{equation}
H_{\rm cav}=\omega_{0}a^{\dagger}a
\end{equation}
is the cavity Hamiltonian, and 
\begin{equation}
H_{I}=g\hat{N}\left(a+a^{\dagger}\right)
\end{equation}
is the junction-cavity interaction.

The equation of motion for the cavity field reads
\begin{align}
\label{eq:mot}
\dot{\tilde{a}}\left(t\right) & =-i\left[\tilde{a}\left(t\right),H\right]\nonumber \\
 & =-i\left(\omega_{0}\tilde{a}\left(t\right)+g\tilde{N}\left(t\right)\right),
\end{align}
where tildes denote operators in the Heisenberg representation ($\tilde{a}\left(t\right)=e^{iHt}ae^{-iHt}$ and $\tilde{N}\left(t\right)=e^{iHt}\hat{N}e^{-iHt}$).
It is convenient to introduce the interaction picture via
\begin{align}
\tilde{N}\left(t\right) & =U^{\dagger}\left(t\right)N^{i}\left(t\right)U\left(t\right),
\end{align}
where
\begin{equation}
U\left(t\right)=\exp\left[-i\int_{-\infty}^{t}dt'H_I^{i}\left(t'\right)\right].
\end{equation}
Assuming weak interactions, we expand
\begin{equation}
\tilde{N}\left(t\right)\approx N^{i}\left(t\right)+i\int_{-\infty}^{t}dt'\left[H_{I}^{i}\left(t'\right),N^{i}\left(t\right)\right].
\end{equation}
Consequently, Eq.~(\ref{eq:mot}) becomes
\begin{align}
\label{eq:mot2}
\dot{\tilde{a}}\left(t\right) & \approx-i\left(\omega_{0}\tilde{a}\left(t\right)+gN^{i}\left(t\right)+gi\int_{-\infty}^{t}dt'\left[H_{I}^{i}\left(t'\right),N^{i}\left(t\right)\right]\right)\nonumber \\
 & =-i\omega_{0}\tilde{a}\left(t\right)-igN^{i}\left(t\right)\nonumber \\
 & +g^{2}\int_{-\infty}^{t}dt'\left(a^{i}\left(t'\right)+a^{i\dagger}\left(t'\right)\right)\left[N^{i}\left(t'\right),N^{i}\left(t\right)\right]\nonumber \\
 & \approx-i\omega_{0}\tilde{a}\left(t\right)-igN^{i}\left(t\right)\nonumber \\
 & +g^{2}\int_{-\infty}^{t}dt'\left(\tilde{a}\left(t'\right)+\tilde{a}^{\dagger}\left(t'\right)\right)\left[N^{i}\left(t'\right),N^{i}\left(t\right)\right],
\end{align}
where we neglected higher order terms in $g$ in the last line.

Next, we write $\tilde{a}\left(t\right)=\tilde{a}_{s}\left(t\right)e^{-i\omega_{0}t}$,
where $\tilde{a}_s\left(t\right)$ evolves
slowly in time. Also, to lowest order in $g$, we replace $\left[N^{i}\left(t'\right),N^{i}\left(t\right)\right]$ by its ground state average. Then, Eq.~(\ref{eq:mot2}) can be approximated as
\begin{align}
\dot{\tilde{a}}_{s}\left(t\right) & \approx-igN^{i}\left(t\right)e^{i\omega_{0}t}\nonumber \\
 & -g^{2}\left[C_{-}\left(t,\varphi\right)\tilde{a}_{s}\left(t\right)+C_{+}\left(t,\varphi\right)\tilde{a}_{s}^{\dagger}\left(t\right)\right],
\end{align}
where
\begin{align}
C_{\pm}\left(t,\varphi\right) & =\int_{-\infty}^{t}dt'e^{i\omega_{0}\left(t\pm t'\right)}\left\langle 0;\varphi\left|\left[N^{i}\left(t\right),N^{i}\left(t'\right)\right]\right|0;\varphi\right\rangle. 
\end{align}

The correlation functions $C_\pm$ can be computed explicitly. First, we consider 

\begin{align}
C_{-}\left(t,\varphi\right)
 & =\int_{0}^{\infty}d\tau\int\frac{d\omega_{1}d\omega_{2}}{\left(2\pi\right)^{2}}e^{i\left(\omega_{0}-\omega_{2}\right)\tau}e^{i\left(\omega_{1}+\omega_{2}\right)t}\nonumber \\
 & \times\left\langle 0,\varphi\left|\left[N^{i}\left(\omega_{1}\right),N^{i}\left(\omega_{2}\right)\right]\right|0,\varphi\right\rangle .
\end{align}
As usual,  one writes
\begin{align}
\int_{0}^{\infty}d\tau e^{i\left(\omega_{0}-\omega_{2}\right)\tau} & =\lim_{\epsilon\to0}\int_{0}^{\infty}d\tau e^{i\left(\omega_{0}-\omega_{2}+i\epsilon\right)\tau}\nonumber \\
 & =P\frac{i}{\left(\omega_{0}-\omega_{2}\right)}+\pi\delta\left(\omega_{0}-\omega_{2}\right).
\end{align}
Therefore,
\begin{align}
\label{eq:b13}
 & C_{-}\left(t,\varphi\right)\nonumber \\
 & =iP\int\frac{d\omega_{1}d\omega_{2}}{\left(2\pi\right)^{2}}\frac{e^{i\left(\omega_{1}+\omega_{2}\right)t}}{\omega_{0}-\omega_{2}}\left\langle 0;\varphi\left|\left[N^{i}\left(\omega_{1}\right),N^{i}\left(\omega_{2}\right)\right]\right|0;\varphi\right\rangle \nonumber \\
 & +\frac{1}{2}\int\frac{d\omega_{1}}{2\pi}e^{i\left(\omega_{1}+\omega_{0}\right)t}\left\langle 0;\varphi\left|\left[N^{i}\left(\omega_{1}\right),N^{i}\left(\omega_0\right)\right]\right|0;\varphi\right\rangle .
\end{align}

Using $\sum_j |j;\varphi\rangle \langle j; \varphi| = {\bf 1}$ and recognizing that 
\begin{align}
\langle 0; \varphi | N^i (t) | j; \varphi\rangle &= e^{-i \left(E_0(\varphi)-E_j(\varphi)\right)t} \langle 0; \varphi | \hat{N} | j; \varphi\rangle \nonumber\\
&\equiv e^{-i \Delta E_{0 j} t} N_{0j},
\end{align}
we obtain
\begin{align}
 & \left\langle 0,\varphi\left|\left[N^{i}\left(t\right),N^{i}\left(t'\right)\right]\right|0,\varphi\right\rangle \nonumber \\
 & =-2i\sum_{j}\sin\left[\Delta E_{0j}\left(\varphi\right)\left(t-t'\right)\right]\left|N_{0j}\left(\varphi\right)\right|^{2}.
\end{align}
Fourier transforming to frequency space, this gives
\begin{align}
\label{eq:b16}
 & \left\langle 0,\varphi\left|\left[N^{i}\left(\omega_{1}\right),N^{i}\left(\omega_{2}\right)\right]\right|0,\varphi\right\rangle \nonumber \\
 & =\left(2\pi\right)^{2}\delta\left(\omega_{1}+\omega_{2}\right)\nonumber \\
 & \times\sum_{j}\left|N_{0j}\left(\varphi\right)\right|^{2}\left[\delta\left(\Delta E_{0j}\left(\varphi\right)-\omega_{1}\right)-\delta\left(\Delta E_{0j}\left(\varphi\right)+\omega_{1}\right)\right].
\end{align}
Substituting Eq.~(\ref{eq:b16}) in Eq.~(\ref{eq:b13}), we get
\begin{align}
 & C_{-}\left(t,\varphi\right)=-2i\sum_{j}\left|N_{0j}\left(\varphi\right)\right|^{2}\frac{\Delta E_{0j}\left(\varphi\right)}{\omega_{0}^{2}-\left(\Delta E_{0j}\left(\varphi\right)\right)^{2}}\nonumber \\
 & +\pi\sum_{j}\left|N_{0j}\left(\varphi\right)\right|^{2} \left[\delta\left(\Delta E_{0j}\left(\varphi\right)+\omega_{0}\right) - (\omega_0\to -\omega_0)\right].
\end{align}
Proceeding similarly for $C_{+}$, we find
\begin{align}
 & C_{+}\left(t,\varphi\right)\nonumber \\
 & =-2ie^{i2\omega_{0}t}\sum_{j}\left|N_{0j}\left(\varphi\right)\right|^{2}\frac{\Delta E_{0j}\left(\varphi\right)}{\omega^{2}-\left(\Delta E_{0j}\left(\varphi\right)\right)^{2}}\nonumber \\
 & +\pi e^{i2\omega_{0}t}\sum_{j}\left|N_{0j}\left(\varphi\right)\right|^{2}\nonumber \\
 & \times\left[\delta\left(\Delta E_{0j}\left(\varphi\right)-\omega_{0}\right)-\delta\left(\Delta E_{0j}\left(\varphi\right)+\omega_{0}\right)\right].
\end{align}
The fact that $C_+(t)$ varies rapidly in time ($\propto e^{2 i \omega_0 t}$) means that it can be discarded in the rotating wave approximation.
We finally obtain, back in the original time frame,
\begin{align}
\dot{\tilde{a}}\left(t\right) & \approx-i\omega_{0}\tilde{a}\left(t\right)-igN^{i}\left(t\right)-g^{2}C_{-}\left(t,\varphi\right)\tilde{a}\left(t\right).
\end{align}
The imaginary part of $C_{-}$ 
renormalizes the cavity resonance frequency, $\omega_0 \to \omega_0 + \bar{\omega}$, where
\begin{equation}
\bar{\omega}=2g^{2}\sum_{j}\left|N_{0j}\left(\varphi\right)\right|^{2}\frac{\Delta E_{0j}\left(\varphi\right)}{\left(\Delta E_{0j}\left(\varphi\right)\right)^{2}-\omega_{0}^{2}}.
\end{equation}
The real part of $C_+$ describes the junction-induced decay of cavity photons, with rate 
\begin{align}
\bar{\kappa} & \equiv 2\pi g^{2}\sum_{j}\left|N_{0j}\left(\varphi\right)\right|^{2}\nonumber \\
 & \times\left[\delta\left(\Delta E_{0j}\left(\varphi\right)+\omega_{0}\right)-\delta\left(\Delta E_{0j}\left(\varphi\right)-\omega_{0}\right)\right].
\end{align}
Therefore,
\begin{equation}
\dot{\tilde{a}}\left(t\right)  \approx -\left(i(\omega_0+\bar{\omega})+\frac{\bar{\kappa}}{2}\right)\tilde{a}\left(t\right)
-ig N^{i}\left(t\right).\label{eq:cav_effB}
\end{equation}
In the presence of input fields, their coupling with the cavity induces an additional damping $\kappa_0$ for the cavity photons, ~\cite{MilburnBook} so that
\begin{equation}
\dot{\tilde{a}}\left(t\right)=-\left(i\omega_R+\frac{\kappa_R}{2}\right)\tilde{a}\left(t\right)
-ig N^{i}\left(t\right)+\sqrt{\kappa_{0}}\tilde{b}_{\rm in}\left(t\right),\label{eq:cav_effB2}
\end{equation}
where $\omega_R=\omega_0+\bar{\omega}$ and $\kappa_R=\kappa_0+\bar{\kappa}$.
This completes the derivation of Eq.~(\ref{eq:cav_eff}) of the main text. 

From Eq.~(\ref{eq:cav_effB2}), one can readily derive Eq.~(\ref{eq:outin}) of the main text.
First, recall that the output field can be related to the input field by the boundary condition
\begin{equation}
\label{eq:bcond}
\tilde{b}_{\rm out}\left(t\right)=\sqrt{\kappa_{0}}\tilde{a}\left(t\right)-\tilde{b}_{\rm in}\left(t\right).
\end{equation}
Combining the Fourier transforms of Eqs.~(\ref{eq:bcond}) and (\ref{eq:cav_effB2}), we obtain
\begin{equation}
\tilde{b}_{\rm out}\left(\omega\right)=\frac{-\left[\omega-\omega_R-i\frac{\kappa_{0}-\bar{\kappa}}{2}\right]\tilde{b}_{\rm in}\left(\omega\right)+g\sqrt{\kappa_{0}}N^{i}\left(\omega\right)}{\omega-\omega_R+i \kappa_R/2}.
\end{equation}
Noting that $\langle 0;\varphi| N^{i}\left(\omega\right)|0; \varphi \rangle \propto\delta\left(\omega\right)$, and recalling that 
we are interested in the response at frequencies close to $\omega_{0}$, we write
\begin{equation}
\langle\tilde{b}_{\rm out}\left(\omega\right)\rangle=\frac{-\left[\omega-\omega_R-i\frac{\kappa_{0}-\bar{\kappa}}{2}\right]\langle\tilde{b}_{\rm in}\left(\omega\right)\rangle}{\omega-\omega_R+i \kappa_R/2}.
\end{equation}
From here, the expressions for the reflection coefficient and phase shift quoted in the main text can be recovered.

\bibliographystyle{apsrev4-1}
\bibliography{bibliog_JJ_int_Cavity}

\end{document}